\documentclass{aa}  
\usepackage{xcolor}

\usepackage{graphicx}

\usepackage{txfonts,amsmath}
\usepackage{bm}
\usepackage{natbib,twoopt}

\usepackage[breaklinks=true,colorlinks=true,
            linkcolor=blue,
            citecolor=blue,
            urlcolor=blue]{hyperref}

\bibpunct{(}{)}{;}{a}{}{,}            
\makeatletter
  \newcommandtwoopt{\citeads}[3][][]{\href{http://adsabs.harvard.edu/abs/#3}
    {\def\hyper@linkstart##1##2{}
     \let\hyper@linkend\@empty\citealp[#1][#2]{#3}}}
  \newcommandtwoopt{\citepads}[3][][]{\href{http://adsabs.harvard.edu/abs/#3}
    {\def\hyper@linkstart##1##2{}
     \let\hyper@linkend\@empty\citep[#1][#2]{#3}}}
  \newcommandtwoopt{\citetads}[3][][]{\href{http://adsabs.harvard.edu/abs/#3}
    {\def\hyper@linkstart##1##2{}
     \let\hyper@linkend\@empty\citet[#1][#2]{#3}}}
  \newcommandtwoopt{\citeyearads}[3][][]
    {\href{http://adsabs.harvard.edu/abs/#3}
    {\def\hyper@linkstart##1##2{}
     \let\hyper@linkend\@empty\citeyear[#1][#2]{#3}}}
\makeatother

\begin{document} 
   \title{Cosmological simulation of a radio synchrotron bridge between pre-merging galaxy clusters} 
   \author{K. Nishiwaki\inst{1},
            G. Brunetti\inst{1},
            F. Vazza\inst{2,1},
            \and
            C. Gheller\inst{1}
          }

   \institute{Istituto di Radioastronomia, INAF, Via Gobetti 101, 40121 Bologna, Italy\,
              \\
              \email{kosuke.nishiwaki@inaf.it}
         \and
             Dipartimento di Fisica e Astronomia, Universita di Bologna, Via Gobetti 92/3, 40121 Bologna, Italy.\\
             }

   \date{Received  ,; accepted , }

  \abstract
   {
   Radio bridges are diffuse synchrotron emission observed between merging galaxy clusters.
   Recent radio observations have reported both detections and non-detections of radio bridges between clusters. The detections imply the presence of cosmic rays (CRs) and magnetic fields permeating the cosmic web that produce synchrotron emission observable with current facilities, whereas the non-detections suggest that specific physical conditions are required for their formation.
   }
   {
   We study the CR reacceleration by solenoidal turbulence in the filament connecting two massive clusters at an early stage of the merger. Our aim is to test whether this mechanism can generate diffuse emission in the inter-cluster region.
   }
   {
   We performed a cosmological magneto-hydrodynamical (MHD) simulation using the Enzo code.
   We improved a run-time Lagrangian tracer method implemented in Enzo, and followed the trajectories of baryonic matter using $N=\mathcal{O}(10^7)$ tracer particles.
   In post-processing, we conducted a parallel computation of the Fokker-Planck (FP) equation for all tracers, with cooling and reacceleration efficiencies evaluated from the local quantities recorded along each tracer trajectory.
   }
   {
   Our simulation generates a megaparsec-sized radio bridge in the early stage of the cluster merger. Within a reasonable parameter range, the reacceleration model produces a broad variety of spectra. In our fiducial model, the simulated bridge matches several properties of the one found between Abell 399 and Abell 401, such as its spectral shape, intensity profile, and pixel-by-pixel correlation between radio and X-ray intensities.
   }
   {
   The inter-cluster region is filled with turbulence induced by infalling mass clumps and subsequently amplified by the approaching motion of the clusters. The CR reacceleration by the turbulence is a viable mechanism to power a megaparsec-sized synchrotron emission observed as radio bridges.
   }


   \titlerunning{cosmological simulation of radio bridge}

   \maketitle

\section{Introduction \label{sec:intro}}
The large-scale structure (LSS) of the Universe is often referred to as a "cosmic web" and it consists of hierarchical parts called nodes, filaments, sheets, and voids \citep[][]{Bond_1996}. 
Numerical simulations predict that $\approx30-40$ \% of the total baryonic matter in the Universe resides in the filaments between galaxy clusters (GCs) in the form of warm-hot intergalactic medium (WHIM) \citep[e.g.,][]{Cen_Ostriker_99,Dave_2001}.
Recently, an increasing number of observations have reported detections of the densest and hottest part of the WHIM between pairs of GCs spaced a few to a few tens of megaparsecs apart \citep[e.g.,][]{Werner_2008,Planck_2013_filaments,Tanimura_2020,Reiprich_2021,Mirakhor_2022,Sarkar_2022,Dietl_2024,Veronica_2024,Migkas_2025}.
The pair of clusters Abell 0399 and 0401 (hereafter A399--A401) is one of the best-studied examples of such inter-cluster filaments \citep[e.g.,][]{Fujita_1996,Fujita_2008,Akamatsu_2017,Hincks_2022,Radiconi_2022}.
It appears to be in an initial stage of the merger \citep[][]{Bonjean_2018} with the projected separation of two clusters of about 3 Mpc.
\par

In the past decades, radio observations have revealed a variety of diffuse synchrotron emission in GCs, such as radio halos and cluster radio shocks \citep[][]{vanWeeren_review}.
The unprecedented sensitivity of recent radio facilities, such as the Low Frequency Array (LOFAR) and MeerKAT, has enabled the detection of diffuse emission on larger scales \citep[e.g.,][]{Rajpurohit_2021,Botteon_2022_A2255,Cuciti_2022,vanWeeren_2024,Rajpurohit_2025}.
\citet{Govoni_2019} reported the first discovery of a ``radio bridge" at 144 MHz in A399--A401, which is diffuse emission covering the region between a pair of merging GCs.

Another radio bridge has been found in the Abell 1758 system \citep[][]{Botteon_2020}, which consists of two massive galaxy clusters.
However, non-detections in other similar systems indicate that such bridge-like emission is not ubiquitous in the filamentary regions between clusters \citep[e.g.,][]{Botteon_2019a,Bruggen_2021,Pignataro_2024}.
The detection of radio bridges suggests that magnetic fields and cosmic-ray electrons (CREs) can permeate into the cosmic web beyond cluster virial radii. 
In contrast, the non-detections indicate that specific physical conditions are required to generate a detectable radio bridge, or that the phenomenon is transient on cosmological timescales.
\par

The mechanisms responsible for magnetic field amplification and particle acceleration in the inter-cluster bridge remain uncertain.
The CREs in the bridge region may be injected through the feedback processes associated with active galactic nuclei (AGNs) or star-forming galaxies \citep[e.g.,][]{Voelk_1996,Berezinsky_1997}.
However, once injected, the CREs are subjected to severe energy losses owing to synchrotron and inverse-Compton radiation, as well as Coulomb collisions.
As a result, the lifetime of radio-emitting CREs ($\sim100$ Myr) is too short for them to propagate over megaparsec scales and to produce smooth, extended bridge emission.
The same problem holds for giant radio halos in GCs and is commonly referred to as the slow-diffusion problem \cite[][]{Brunetti_Jones_review}.
Reacceleration of CREs is therefore often invoked as a solution to this problem in the cluster environments \citep[e.g.,][]{Brunetti_2001,Peterson_2001,Fujita_2003}.
\par

Numerical simulations have shown that the intra-cluster medium (ICM) and inter-cluster bridges are permeated by shocks and turbulence driven by continuous matter accretion and episodic merger events of dark matter halos \citep[e.g.,][]{Ryu_2003,Pfrommer_2006,Nelson_2014,Steinwandel_2024}.
Stochastic Fermi acceleration at these shocks or within turbulent flows may not be efficient enough to accelerate CREs directly from the thermal pool, but it can counteract radiative energy loss of the preexisting CREs \citep[e.g.,][]{Brunetti_Jones_review}.
Secondary electrons generated through the hadronic interaction of CR protons can also contribute to diffuse synchrotron emission \citep[e.g.,][]{Dennison_1980,Blasi_Colafrancesco_99}.
However, in low-density gas regions such as cluster outskirts, the contribution from secondary electrons is expected to be smaller than in the central regions of clusters.
\par

Using a cosmological magnetohydrodynamical (MHD) simulation, \citet{Govoni_2019} suggest that weak shocks with Mach numbers of $M \approx 2$–3 in the region of the radio bridge could reaccelerate CREs and produce apparently smooth and volume-filling emission under particularly favorable conditions.
Their shock reacceleration model requires the seed population of CREs to fill most of the bridge volume, while radiative energy losses limit their lifetimes to be less than 1 Gyr.
This requirement implies a degree of fine-tuning and is hard to reconcile with the dynamical mixing time of the 3 Mpc-long bridge region, which should be longer than the sound-crossing time ($\gtrsim$ 3 Gyr) \citep[][]{Govoni_2019}.

Using a snapshot from the same simulation, \citet{BV20} explored an alternative scenario involving the turbulent reacceleration by the solenoidal component of the turbulence \citep[][]{BL16}.
Although they relied on a simplified single-zone model, they succeeded at reproducing the radio spectrum and showed that the projected emission is dominated by strongly turbulent regions with a relatively small volume-filling factor of approximately 15\%.
Taking into account the spatial distribution of the turbulent energy and the line-of-sight projection effect, \citet{Nishiwaki24} show that the steep-spectrum emission in the cluster outskirts can be reproduced when approximately 1\% of the turbulent energy is channeled into the CR reacceleration.
\par

The aim of this paper is to explore the reacceleration model while overcoming the limitations arising from the simplifications adopted in previous studies.
In particular, we investigate the coupled time evolution of thermal and nonthermal components in a filament between two GCs using advanced numerical simulations.
In these simulations, we explicitly followed the transport and temporal evolution of CREs.
We include a variety of physical processes governing CRE energy gains and losses, such as radiative and collisional cooling, adiabatic expansion and compression, and turbulent reacceleration.

The efficiencies of these processes depend on local turbulence properties, magnetic field strength, and gas density.
To study the evolution of CR spectrum during a cluster merger, it is therefore necessary to consider the evolution of the background fields along the trajectories of CREs.
This requirement motivates us to perform parallel simulations of the Fokker-Planck (FP) equation for a variety of trajectories.
\par

Several previous works attempted to model energy-resolved CR distributions in GCs and LSS by combining cosmological MHD simulations with FP solvers. While earlier studies relied on MHD snapshots in post-processing \citep[e.g.,][]{Donnert_Brunetti_2014,Pinzke_2017,Winner_2019,Smolinski_2023,Beduzzi_2024}, more recent frameworks implement on-the-fly FP solvers under approximations that reduce the number of energy or momentum bins \citep[e.g.,][]{Hopkins_2022,Boess_2023}.
We extend these approaches by incorporating the tracer-particle code Crater \citep{Wittor_2017} into the Enzo framework, and performing FP simulations in post-processing. The tracer particles generated and propagated during the runtime of Enzo better represent the baryonic flow compared to post-processing approaches such as Crater.
We apply the FP solver developed in \citet{Nishiwaki24,Nishiwaki25}. Tests of the tracer and FP methods are presented in Appendix.
\par

This paper is structured as follows: In Sect.~\ref{sec:method}, we introduce the set-up of our Enzo simulation and our tracer particle method, and explain the model for the FP simulation.
In Sect.~\ref{sec:results}, we show the results of our simulation, focusing on synchrotron spectrum, one-dimensional profiles of turbulence and emission, and pixel-by-pixel correlation between thermal and nonthermal emission.
In Sect.~\ref{sec:discussion}, we discuss how the results are affected by the parameters and clarify the limitations of our model. The implementation of the tracer method in Enzo is detailed in Appendix. 
Finally, we summarize our conclusions in Sect.~\ref{sec:conclusion}.
In the following sections, we assume a flat $\Lambda$CDM cosmology with $H_0 = 72$ km s$^{-1}$ Mpc$^{-1}$, $\Omega_{\rm m} = 0.258$, $\Omega_{\rm \Lambda} = 0.742$, and $\sigma_8 = 0.8$.

\begin{figure*}
    \centering
    \includegraphics[width=\linewidth]{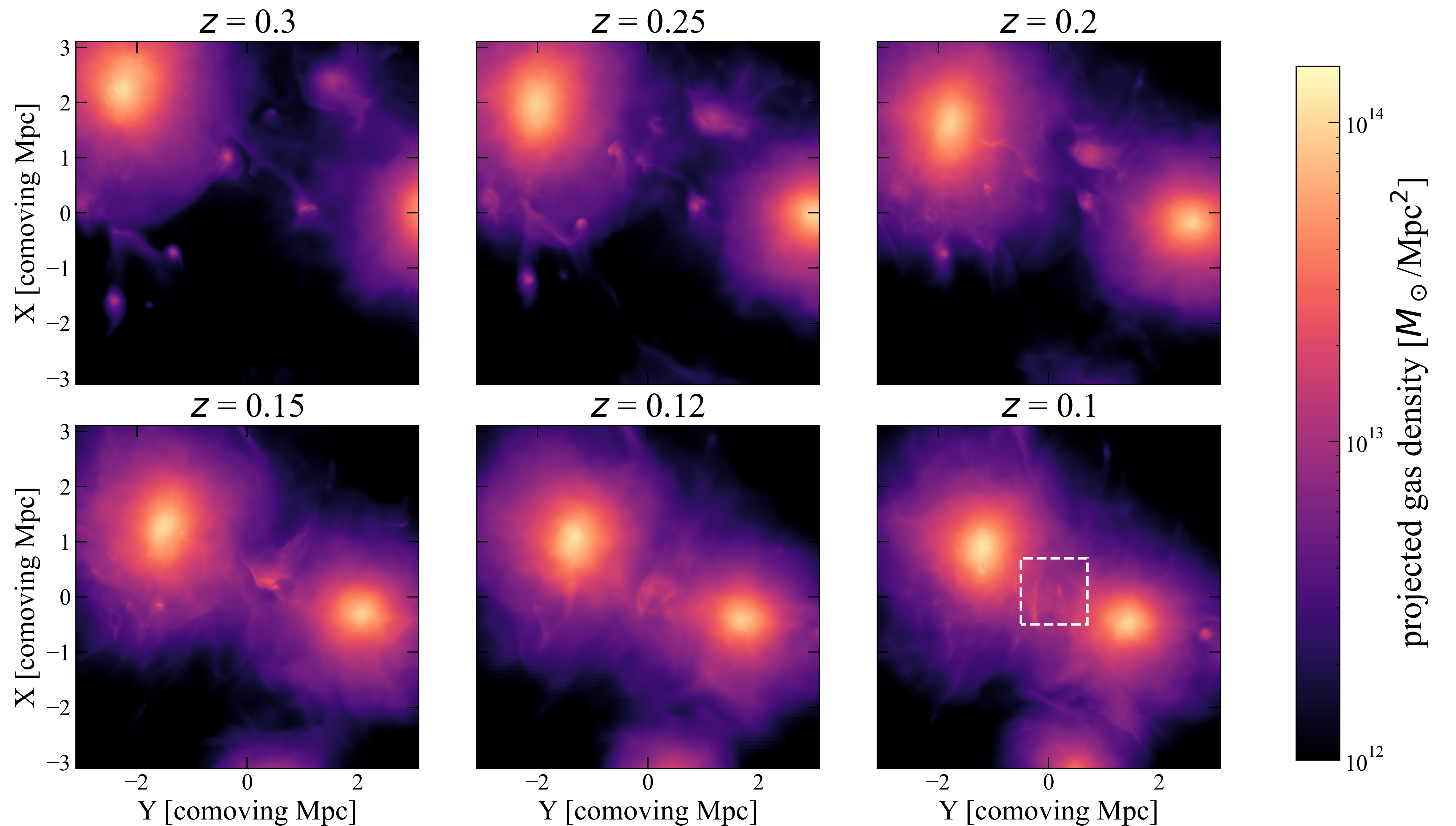}
    \caption{
    Evolution of the projected gas density map along the Z axis in our Enzo simulation. The maps of the zoomed-in region centered on the inter-cluster filament are shown. 
    The white square shows the (1.2 Mpc)$^3$ box region where a radio bridge forms (Sect.~\ref{sec:bridge}).
    }
    \label{fig:Enzo_map_z}
\end{figure*}

\section{Method\label{sec:method}}
In Sect.~\ref{sec:Enzo}, we present the set-up of the Enzo simulation of the merging clusters. During the run of the simulation, $N\approx 1.5\times 10^7$ tracer particles are generated to sample the baryonic fluid elements. In Sect.~\ref{sec:tracer}, we explain the methods for tracer particle generation and propagation in the Enzo simulation.
The strength of solenoidal turbulence around each tracer particle is quantified using the local vorticity (Sect.~\ref{sec:sim_turb}).
The magnetic field in our Enzo simulation may lead to an underestimation of magnetic-field amplification by the turbulent dynamo due to the limited spatial resolution as well as partially due to the use of the Dedner hyperbolic divergence-cleaning scheme to damp numerical errors in the divergence of the magnetic field.
For this reason, we estimate a dynamo-amplified field in post-processing (Sect.~\ref{sec:dynamo}) and adopt the larger of the simulated and modeled values.
In Sect.~\ref{sec:FP}, we introduce the FP equation adopted in this work and summarize the model parameters.

\subsection{Enzo simulation\label{sec:Enzo}}
We simulate a pair of merging galaxy clusters by running the cosmological MHD code Enzo \citep[][]{Bryan_2014}. We perform a new re-simulation of the same system studied in \citet{Govoni_2019} to make use of our developed tracer particle method on board Enzo.
The simulation box is centered on the inter-cluster filament between merging clusters, resembling the A399--A401 system at low redshift. 
The root grid covers a comoving volume of $(260~\mathrm{Mpc})^3$, resolved with $220^3$ cells. 
Following \citet{Govoni_2019}, we employ the MHD solver with Dedner divergence cleaning \citep[][]{Dedner_2002}. We set a uniform comoving magnetic field with components $(B_x,B_y,B_z)=(10^{-10},\,10^{-10},\,10^{-10})~\mathrm{G}$ at the initial redshift $z = 30$ of the simulation. The radiative cooling of the baryonic gas is not included in this simulation.

We employ the adaptive mesh refinement (AMR) in the innermost (30 Mpc)$^3$ region centered on the inter-cluster filament.
Refined grids are introduced in the regions where the mass density of the baryonic gas or dark matter particles exceeds $0.1\rho_{\rm m}$, where $\rho_{\rm m}$ denotes the mean mass density of the Universe.
The cell width is reduced by a factor of two at each refinement level.
The maximum AMR level is eight, which corresponds to 4.2 comoving kpc per cell.
From $z=2$, we decrease the refinement threshold to $0.01\rho_{\rm m}$ and introduce a super-Lagrangian level dependence of the refinement threshold with the exponent of $-0.8$.
With this prescription, the refinement threshold decreases by a factor of $2^{-0.8} \simeq 0.57$ at each successive AMR level.
These aggressive criteria ensure that most of the bridge volume with $\rho_{\rm gas}\gtrsim10^{-28}$ g cm$^{-3}$ is resolved with the resolution of AMR level seven (8.4 comoving kpc) in $z \lesssim 0.15$.
\par

The resolution is sufficiently high to study the bridge region with a resolution of several tens of kiloparsecs, which is comparable to or smaller than the spatial resolution at which the radio bridges are detected.
However, it is not sufficient to resolve the magnetic-field amplification through turbulent dynamo because the dominant scale of the turbulent dynamo in ICM would be $\sim$0.1-1 kpc (Sect.~\ref{sec:dynamo}).
As a result, the simulated magnetic field can be systematically underestimated.
We circumvent this problem by adopting a post-processing model for the magnetic field constrained by the previous MHD simulations dedicated to the study of the turbulent dynamo \citep[e.g.,][]{Beresnyak_Miniati_2016,Vazza_2018_dynamo}.
The dynamo model is explained in Sect.~\ref{sec:dynamo}.
\par

The projected distance between the centers of the two clusters along the $x$-axis of the simulation becomes $\approx 3$ Mpc at $z\approx0.1$, and the simulated system resembles A399--A401.
In Fig.~\ref{fig:Enzo_map_z}, we show the projected gas density map of the bridge region. 
In addition to the two main clusters, numerous smaller clumps move around the clusters and along the filament.
Not only the major merger between the two clusters, but also the collisions of those small clumps act as important sources of the turbulence in the filament (Sect.~\ref{sec:results}).
The white square highlights the region where diffuse emission resembling a radio bridge emerges in the simulation.
In Sect.~\ref{sec:bridge}, we extract data from this region, corresponding to a volume of (1.2 Mpc)$^3$ and study the evolution of turbulence and emission.
\par

\begin{figure*}
    \begin{minipage}{0.48\linewidth}
        \includegraphics[width=\linewidth]{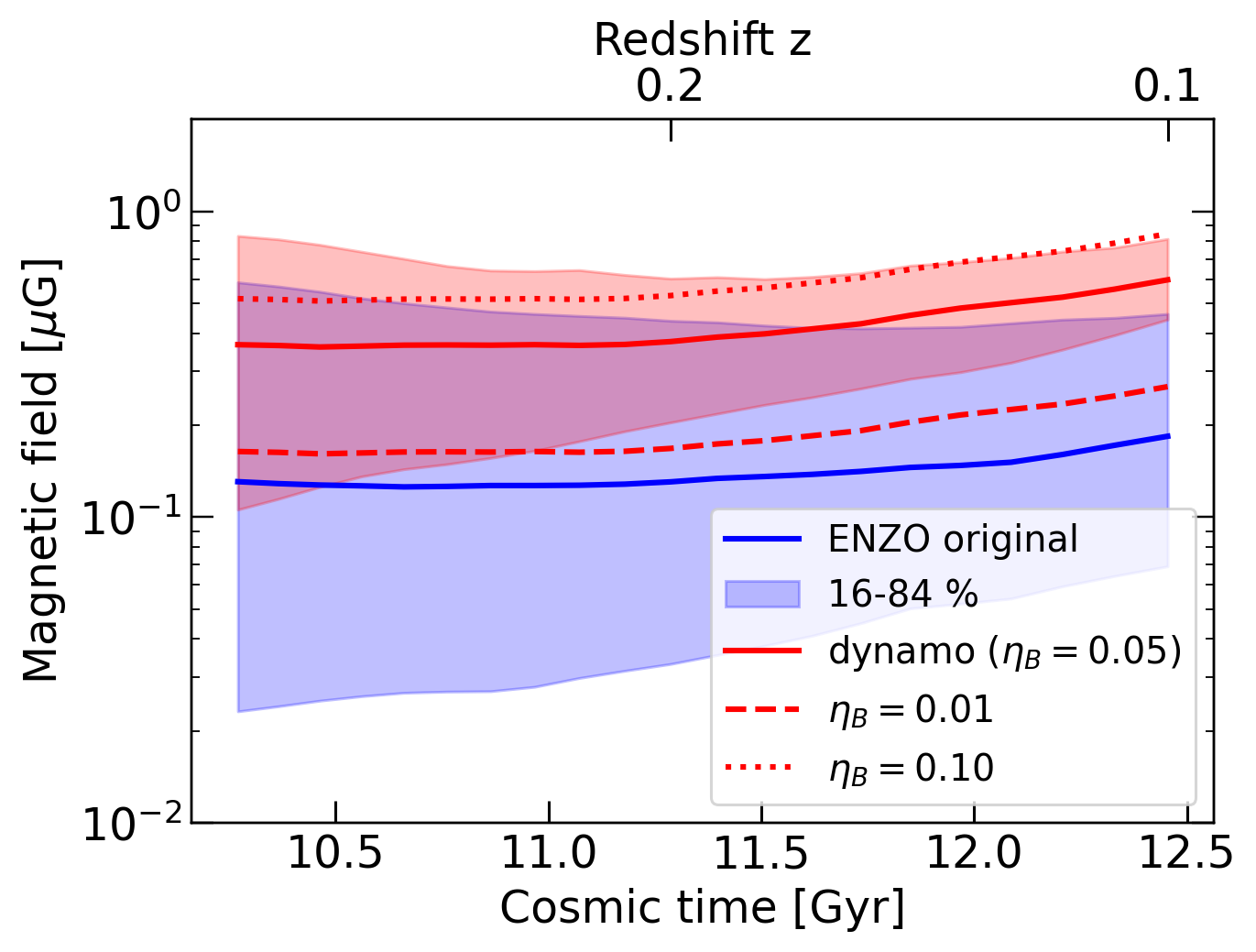}
    \end{minipage}
    \begin{minipage}{0.48\linewidth}
        \includegraphics[width=\linewidth]{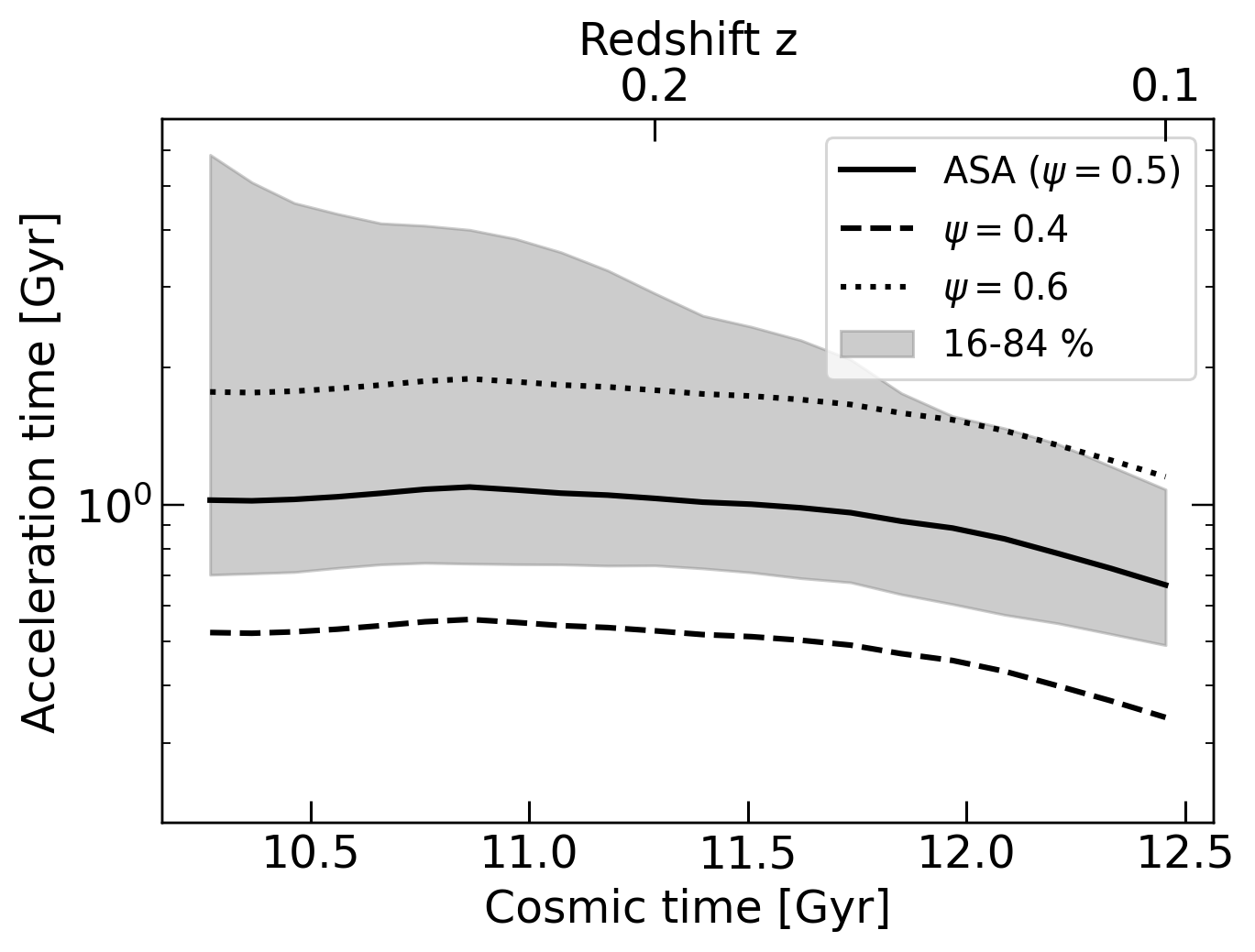}
    \end{minipage}
    \caption{
    Time evolution of the magnetic field (left panel) and timescale of reacceleration (right panel) from $z=0.3$ to 0.1 along the trajectories of $N=5.1\times10^5$ tracer particles that end up in the bridge region (white box in Fig.~\ref{fig:Enzo_map_z}).
    In both panels, the solid line and the shaded region show the median value and 1$\sigma$ range of the distribution.
    In the left panel, the red and blue lines show the dynamo magnetic field calculated with $\eta_B = 0.05$ and the original magnetic field in the Enzo simulation, respectively. The dashed and dotted lines show the median values for different parameters.
    }
    \label{fig:B_tacc_evo}
\end{figure*}

\subsection{Lagrangian tracers \label{sec:tracer}}
In many previous studies, the generation and propagation of Lagrangian tracer particles have been done in post-processing \citep[e.g.,][]{Vazza_2010,Wittor_2017,Beduzzi_2023,Smolinski_2023}. 
In this approach, particle positions are updated according to the local fluid velocity obtained by interpolating grid data from two consecutive simulation snapshots. 
However, because the time interval between snapshots is generally much larger than the hydrodynamical or MHD time step at high AMR levels, this method can lead to systematic over- or undershooting of tracer particle positions.

We improve the accuracy of the tracers by generating them and following their trajectories on the fly during the run of the cosmological MHD simulation. To this end, we extend the tracer algorithm natively available in Enzo.
Here, we briefly summarize our tracer generation and propagation methodology; further details are provided in Appendix~\ref{app:trc_generation}.
\par

During the run of Enzo, the tracer particles are generated at a single epoch at $z = 1$ within a comoving (20 Mpc)$^3$ region encompassing the pair of clusters of interest.
The total mass in the region, $M^{\rm tot}_{\rm trc} \approx 3\times10^{14}$ M$_\odot$, is sampled by tracer particles with masses in the range $[M_{\rm trc}^{\rm min},\, M_{\rm trc}^{\rm max}]$.
Below, we briefly overview the tracer particle generation strategy; further details are provided in Appendix~\ref{app:trc_generation}.
\par

We iterate over all grid cells within the $(20~\mathrm{Mpc})^3$ region and evaluate the baryonic mass of each cell, $M_{\rm cell}$. 
If $M_{\rm cell}$ lies within the range $[M_{\rm trc}^{\rm min},\, M_{\rm trc}^{\rm max}]$, the cell mass is represented by a single tracer particle with $M_{\rm trc}=M_{\rm cell}$. 
When $M_{\rm cell} > M_{\rm trc}^{\rm max}$, which typically occurs at coarse resolution, the mass is represented by $\left\lfloor M_{\rm cell}/M_{\rm trc}^{\rm max} \right\rfloor$ tracer particles with $M_{\rm trc}=M_{\rm trc}^{\rm max}$, with one additional particle accounting for the residual mass. 
When $M_{\rm cell} < M_{\rm trc}^{\rm min}$, the cell mass is sampled probabilistically by generating a tracer particle with mass $M_{\rm trc}^{\rm min}$ with a probability of $M_{\rm cell}/M_{\rm trc}^{\rm min}$.
\par

In total, we generate $N \approx 1.5\times10^7$ particles at $z=1$.
More than $90\%$ of the tracer particles have masses $M_{\rm trc} < 3\times10^7$ M$_\odot$ (see Appendix~\ref{app:trc_generation}).
As discussed in Sect.~\ref{sec:limitation}, generating tracers only once during the simulation is a limitation of this work because it does not account for CRs freshly injected by AGN, or by shocks,  at later epochs.

After tracer generation, we update tracer positions by assigning the tracer velocity to the local gas velocity, $\bm{u}(\bm{x}) = \bm{v}(\bm{x})$, where the gas velocity at the tracer position is estimated using cloud-in-cell (CIC) interpolation.
However, as pointed out in previous studies \citep[e.g.,][]{Genel_2019}, this procedure can induce artificial clustering of tracer particles in high-density regions, particularly in the presence of turbulence and shocks (see also Appendix~\ref{app:trc_map}).
Following \citet{Wittor_2017}, we alleviate this effect by introducing a perturbation term that models local turbulent mixing,
\begin{equation}
    \delta \bm{v}_{i,j,k}(\bm{x}) = \bm{v}_{i,j,k} - \frac{\sum_{i = 0}^{2}\sum_{j = 0}^{2}\sum_{k = 0}^{2} \bm{v}_{i-1,j-1,k-1}}{27},
    \label{eq:delta_v_trc}
\end{equation}
where $(i,j,k)$ represent the indices of the nearest grid cell of the tracer at the position $\bm{x}$.
We add this term to the tracer velocity, such that $\bm{u}(\bm{x}) = \bm{v}(\bm{x}) + \delta \bm{v}_{i,j,k}$. 
In addition, we upgrade the tracer time integration scheme to a second-order Runge--Kutta method, whereas the original Enzo~2.6 implementation employs a first-order scheme. 
We compare the run-time tracer method used in this work with the post-processing tracer approach in Appendix~\ref{app:trc_map}.
The run-time tracer scheme reproduces the underlying baryonic matter distribution in the Enzo simulation within a relative error of $\lesssim 0.3$ in the region of interest (ROI), while reducing artificial small-scale structures compared to the post-processing method.
Since we assume that the CR population is proportional to the gas mass (Sect.~\ref{sec:seed}), the run-time tracer method has the advantage of reducing small-scale sampling artifacts in the modeled radio emission.
\par

Each tracer particle records its position and local MHD quantities, including density, temperature, magnetic field strength, and velocity, interpolated with CIC.
The interpolation always employs the finest AMR grid available at the tracer position.
Tracer outputs are written at intervals of $\Delta z = 0.01$ over the redshift range $0 < z < 1$, and these data are used as inputs for the Fokker–Planck solver (Sect.~\ref{sec:FP}).

\subsection{Measuring the local turbulent field \label{sec:sim_turb}}
Solenoidal turbulence may play an important role in both magnetic-field amplification by the turbulent dynamo and CR reacceleration in the ICM \citep[e.g.,][]{Cho_Vishniac_2000,BL16,Vazza_2018_dynamo,Nishiwaki24}.
Although the compressive component of turbulence may also contribute to CR acceleration \citep[e.g.,][]{BL07}, we focus exclusively on the solenoidal mode in this work and defer a detailed comparison between the two modes to future studies. 
The local solenoidal turbulent velocity is estimated from the vorticity using the relation $\delta v(l) \approx l|\nabla\times \bm{v}|$, where $l$ denotes the spatial scale at which the vorticity is evaluated. 
We calculate $\nabla\times\bm{v}$ using the central-difference scheme applied to the velocity field on a stencil of $3^3$ cells surrounding the nearest grid cell of each tracer particle.
With this approach, tracer particles probe turbulent motions on different scales, depending on the resolution of their host AMR grid.
For the calculation of the dynamo-amplified magnetic field and the CR acceleration efficiency in the following sections, we normalize the turbulent velocity $\delta v$ to a reference scale of $l = L = 150$ comoving kpc, assuming a Kolmogorov scaling relation $\delta v(l) \propto l^{1/3}$.
\par

The amplitude of solenoidal turbulence can also be measured by separating bulk flows from turbulent motions using filtering techniques and Hodge--Helmholtz decomposition \citep[e.g.,][]{Vazza_2012,Valles-Perez_2024}. 
Turbulence inferred from the local vorticity is qualitatively consistent with that obtained using filtering and decomposition methods \citep[e.g.,][]{Porter_2015,Vazza_2017,Dominguez-F_2024}.
Meanwhile, local vorticity measurements can be biased toward large values, especially in the presence of shock-like discontinuities.
Following previous studies \citep[e.g.,][]{Beduzzi_2024}, we impose an upper limit on the solenoidal turbulent velocity as $M_{\rm s} \equiv \delta v(L)/c_s \leq 0.5$ in order to avoid overestimating $\delta v$ and to keep the CR reacceleration efficiency at a conservative level.

\subsection{Magnetic-field amplification through turbulent dynamo \label{sec:dynamo}}
The turbulent dynamo is one of the most plausible mechanisms for the amplification of magnetic fields in galaxy clusters. 
Gas in the inter-cluster bridge has densities and temperatures similar to those of the cluster outskirts ($n_e\sim10^{-4}$ cm$^{-3}$ and $T\sim 3$ keV), and thus constitutes a weakly collisional high-beta plasma.
Under these conditions, microphysical instabilities can suppress the effective viscosity and allow super-Alfv\'enic turbulence to cascade down to small scales \citep[e.g.,][]{Schekochihin_2006, BL07,Beresnyak_Miniati_2016,Brunetti_Jones_review,Kunz_2022}.  
However, resolving the turbulent dynamo directly in numerical simulations is computationally very demanding. 
The characteristic scale of this is the so-called Alfv\'en scale, 
\begin{equation}
    l_{\rm A} = L \left( \frac{\delta v(L)}{v_{\rm A}} \right)^{-3},
    \label{eq:l_A}
\end{equation}
where $v_{\rm A}$ denotes the Alfv\'en speed.
For turbulence in the ICM, the Alfv\'en Mach number is $M_{\rm A}\equiv\delta v(L)/v_{\rm A} \approx 10$, implying that $l_{\rm A}$ is three or four orders of magnitude smaller than the driving scale of turbulence in GCs ($\sim 0.1-1~{\rm Mpc}$).
\par

Unlike some previous studies that had a high enough spatial resolution to resolve the turbulent dynamo at the cluster centers \citep[e.g.,][]{Vazza_2018_dynamo,Steinwandel_2024}, the finest resolution of our simulation (4.2 kpc) is coarser than $l_{\rm A}$ throughout most of the simulated volume.
For this reason, we use a sub-grid model to evaluate the magnetic field strength in post-processing, following earlier works \citep[][]{Ryu_2008,BV20,Nishiwaki24,Beduzzi_2024}. 
We adopt the relation of ${B_{\rm dyn}^2}/8\pi\sim \eta_BF_{\rm turb}t_{\rm eddy}\sim1/2\eta_{B}\rho\delta v^2$, where $\eta_B$ is the dynamo efficiency, $F_{\rm turb} = 1/2\rho\delta v^3(L)/L$ is the turbulent energy flux, and $t_{\rm eddy} = L/\delta v(L)$ is the eddy turn-over time.
In Kolmogorov turbulence, $F_{\rm turb}$ is independent of the eddy size.
The value of $\eta_B$ remains uncertain; previous numerical simulations suggest $\eta_B \sim 0.01$–$0.1$ \citep[e.g.,][]{Beresnyak_Miniati_2016,Vazza_2018_dynamo}. In our fiducial model, we adopt $\eta_B = 0.05$.
\par

For each tracer particle at each redshift, we compare the magnetic-field strength recorded in the tracer data, $B$, with the dynamo-amplified field $B_{\rm dyn}$ estimated using the method described above, and we adopt the larger of the two for the FP simulation. 
When estimating $B_{\rm dyn}$, we impose the upper limit on $\delta v$ discussed in Sect.~\ref{sec:sim_turb}.
We note that $B_{\rm dyn}$ depends not only on the dynamo efficiency $\eta_B$ but also on the reference scale $L$ at which the turbulent velocity $\delta v$ is measured. 
Our fiducial choice of the parameters, $(L,\eta_B) = (150~{\rm kpc},0.05)$, yields a magnetic-field strength of a few $\mu$G in the central regions of the clusters, consistent with simulations with resolved dynamo and with observational constraints from Faraday rotation measures \citep[e.g.,][]{Bonafede_2010}.
In the following analysis, we vary $\eta_B$ while keeping $L = 150$~kpc fixed.
Models with larger and smaller values of $\eta_B$ are explored in Sect. ~\ref{sec:models}.
\par

We select tracer particles located in the bridge region at $z = 0.1$ and compare the time evolution of $B_{\rm dyn}$ estimated with the dynamo model and that of the magnetic field in the original Enzo simulation, as shown in the left panel of Fig.~\ref{fig:B_tacc_evo}.
The median magnetic field estimated in the post-processing is typically larger than the simulated field by a factor of $\approx$3. 
Within the uncertainty associated with $\eta_B$, the magnetic-field strength in the bridge region spans the range 0.2$\mu$G $\lesssim B_{\rm dyn} \lesssim$ 0.8$\mu$G.
The plasma beta in the inter-cluster bridge ($\rho_{\rm gas}\approx2\times10^{-28}$ g cm$^{-3}$ and $T\approx 4$ keV) remains as high as $\beta_{\rm pl} \approx 50$ even for $\eta_B = 0.1$.
Because the magnetic field is dynamically subdominant in this high-beta regime, the adopted dynamo model does not introduce significant tension with the fluid dynamics of the original Enzo simulation.

\subsection{Fokker--Planck simulation and model parameters \label{sec:FP}}
\subsubsection{Fokker--Planck equation and the solver \label{sec:FP_eq}}
For all tracer particles in the simulation, we solve the following Fokker--Planck (FP) equation to study the evolution of the spectral momentum distribution of CREs, $N(p)$,
\begin{equation}
    \frac{\partial N(p)}{\partial t} = \frac{\partial}{\partial p}\left[D_{pp}\frac{\partial N(p)}{\partial p}+\left(b(p,z(t))-\frac{2}{p}D_{pp}\right)N(p) \right] + Q(p,z(t)),
    \label{eq:FP}
\end{equation}
where $b$, $D_{pp}$, and $Q$ denote cooling, momentum diffusion, and injection coefficients, respectively. 
The cooling term includes radiative losses due to synchrotron and inverse-Compton emission, Coulomb cooling, and adiabatic compression or expansion \citep[see][for the detail]{Nishiwaki25}.
In this work, we consider CR injection at a single epoch $z = z_{\rm i}$ and set $Q(p)\equiv0$ at all other times. 
For simplicity, we neglect the spatial diffusion of CREs, injection via diffusive shock acceleration, and the production of secondary electrons though hadronic interactions between CR protons and thermal protons. We instead focus on the effect of turbulent reacceleration in this work.
\par

We set 128 momentum bins that are equally spaced in logarithmic scale over the range $p_{\rm min}\leq p \leq p_{\rm max}$. We set $p_{\rm min}/m_ec = 0.1$ and $p_{\rm max}/m_ec  = 10^6$, where $m_e$ denotes the electron mass. 
Following our previous work, we adapt the fully implicit Chang-Cooper scheme to numerically integrate Eq.~(\ref{eq:FP}) \citep[][]{Chang_Cooper_70,Park_Petrosian_96}.

This scheme has also been applied in the previous papers in the field \citep[e.g.,][]{Donnert_Brunetti_2014,Vazza_2023}.
In Appendix~\ref{app:FP}, we assess the accuracy of the FP solver by performing test problems with known analytical solutions.
\par

The timestep adopted for the FP solver is shorter than the tracer snapshot interval, $\Delta T_{\rm snp}$.
We set the FP timestep to the minimum of the cooling time of electrons with $p = p_{\rm max}$ and $\Delta T_{\rm snp}/400$.
When a tracer encounters a strongly turbulent eddy, the local eddy turnover time ($t_{\rm eddy} = L/\delta v$) can be shorter than $\Delta T_{\rm snp}$.
Because reacceleration by such an eddy is not expected to persist much longer than $t_{\rm eddy}$, we limit the effective reacceleration duration to $\approx2t_{\rm eddy}$. 
In practice, we apply $D_{pp}$ computed from Eq.~(\ref{eq:Dpp}) for $rN_{\rm sblp}$ substeps and $D_{pp} = 0$ for $(1-r)N_{\rm sblp}$ substeps, where $N_{\rm sblp}$ is the number of FP substeps within one snapshot interval and $r = 2t_{\rm eddy}/\Delta T_{\rm snp}$. The FP solver is parallelized using MPI, as each tracer particle evolves independently.
\par

\subsubsection{Reacceleration model \label{sec:reacc}}

We adopt the model proposed by \citet{BL16} for the stochastic acceleration of nonthermal particles in super-Alfv\'enic turbulence. 
In super-Alfv\'enic turbulence the magnetic-field lines are tangled on a minimum scale on the order of the Alfv\'en scale, $l_{\rm A}$. 
This scale sets the relevant size of the regions where magnetic reconnection and dynamo occur within the turbulent cascade.
Within reconnection regions particles interacting with converging magnetic-field lines are accelerated \citep[e.g.,][]{deGouveiadalPino_Lazarian_2005,Drake_2006,Drury_2012,Kowal_2012,Guo_2024}, whereas energetic particles within stretching and expanding magnetic-field lines in dynamo regions are expected to cool.
\citet{BL16} have modeled the stochastic reacceleration process due to repeated encounters of energetic particles diffusing across reconnection and dynamo regions. 
The process results in a mechanism that is similar to a second-order Fermi mechanism with a particles momentum-diffusion coefficient,

\begin{equation}
    D_{pp} = 3 \sqrt{\frac{5}{6}}\frac{c_s^3}{c}\frac{\sqrt{\beta_{\rm pl}}}{L}M_{\rm s}^3\psi^{-3}p^2,
    \label{eq:Dpp}
\end{equation}
where $\beta_{\rm pl}$ denotes the plasma beta, and $\psi$ represents the CR particle mean free path normalized to the Alfv\'en scale $l_{\rm A}$.
We treat $\psi$ as a free model parameter.
We note that $D_{pp}$ can depend on the dynamo efficiency $\eta_B$ through its dependence on the magnetic-field strength.
When $B = B_{\rm dyn}$, the momentum-diffusion coefficient scales as $D_{pp} \propto \eta_B^{-1/2}$. 
In \citet{Nishiwaki24}, we explore the parameter space of $\eta_B$ and $\psi$ in modeling synchrotron emission in the peripheral regions of galaxy clusters (see their Fig.~3).
Motivated by the steep radio spectrum of the bridge emission around 100~MHz \citep[e.g.,][]{Pignataro_2024}, we adopt $\eta_B\approx 0.01 -0.1$, $\psi \approx 0.2 - 0.6$ as reference ranges of the parameters.
In our fiducial model, we use $\psi = 0.5$.
\par

In the right panel of Fig.~\ref{fig:B_tacc_evo}, we show the evolution of the reacceleration timescale, defined as $t_{\rm acc} = p^2/4D_{pp}$, for the same particle population as in the left panel.
The imposed upper limit on the turbulent Mach number, $M_{\rm s} \leq0.5$ (Sect.~\ref{sec:sim_turb}), leads to an asymmetric distribution of $t_{\rm acc}$.
The median reacceleration timescale gradually becomes smaller as the turbulence is amplified (Sect.~\ref{sec:results}) and reaches $t_{\rm acc}\approx 600$ Myr at $z = 0.1$.
The dashed and dotted curves show the strong dependence of the acceleration timescale on the parameter $\psi$ for a fixed value of $\eta_B = 0.05$.
As discussed in Sect.~\ref{sec:models}, the resulting synchrotron spectrum is significantly affected by the choice of  $\psi$.
\par

\begin{table}[]
    \centering
    \caption{Parameters in our FP simulation.}
    \begin{tabular}{ccc}
      \hline
      \hline
       $L$  & reference scale for turbulence &150 kpc \\
    $\eta_B$ & dynamo efficiency & 0.05 \\
    $z_i$  & redshift of CR injection &1 \\
    $\phi$  & initial CREs to thermal electrons & $6\times10^{-8}$ \\
    $\rho_{\rm thr}$ & threshold density for the CR injection & 10$^{-28}$ g cm$^{-3}$ \\
    $\psi$ & mean free path of CREs relative to $l_{\rm A}$ &0.5 \\
     \hline
    \end{tabular}
    \label{tab:params}
\end{table}

\subsubsection{Initial cosmic-ray distribution \label{sec:seed}}
As an initial condition, we set a single power-law spectrum of $N(p|z_{\rm i}) = N_0p^{-\delta}$ with spectral index $\delta = 2.2$ at redshift $z = z_{\rm i}$. 
The minimum injected momentum of CREs is set to $p_{\rm min}^{\rm inj} = 10$.
To determine the normalization $N_0$, we assume a fixed ratio $\phi$ between the number of the injected CREs and that of the thermal electrons associated with each tracer particle \citep[][]{Beduzzi_2024},
\begin{equation}
    \phi = \frac{N_{\rm CRE}^{\rm inj}}{N_{\rm th,e}},
    \label{eq:phi}
\end{equation}
where $N_{\rm CRE}^{\rm inj} = \int_{p_{\rm min}^{\rm inj}}^{p_{\rm max}}N_0p^{-\delta}dp$, $N_{\rm th,e} \approx 0.52M_{\rm trc}/(\mu_{\rm mol}m_p)$, and $\mu_{\rm mol}\approx0.6$ is the mean molecular weight.
Due to the variation in the tracer mass $M_{\rm trc}$, the normalization $N_0$ varies for a fixed value of $\phi$. 

This is an additional simplification of the model. It is motivated by the physical expectation that CR sources, such as relativistic AGN jets and galactic winds from star-forming galaxies, are more numerous in overdense regions.
This expectation is qualitatively supported by recent simulation by \citet{Vazza_2025}, who considered CR seeding by cosmological shocks as well as feedback from AGNs and galactic winds and found that the large-scale CR distribution in the cosmic web broadly traces the gas distribution.

\citet{Beduzzi_2024} adopted the scaling of Eq.~(\ref{eq:phi}) and reproduced the brightness profiles of radio mega-halos with $\phi \sim 10^{-6}$. 
Note that the value of $\phi$ depends on $\delta$ and $p_{\rm min}^{\rm inj}$, which remain poorly constrained by observations.
\par

The above seeding prescription may not be applicable in low-density regions where the CR sources are sparsely distributed.
Assuming that CR seeding operates predominantly in the high density parts of the cosmic web, we introduce a threshold in the initial gas density, $\rho_{\rm thr}$.
In practice, we solve the FP equation for all tracer particles and set $N_0 = 0$ in post-processing for tracers whose initial gas density satisfies $\rho < \rho_{\rm thr}$. 
We treat $z_{\rm i}$, $\phi$, and $\rho_{\rm thr}$ as the model parameters.
In Sect.~\ref{sec:results}, we adopt a fiducial model with $z_{\rm i} = 1$ and $\rho_{\rm thr} = 10^{-28}$ g cm$^{-3}$ ($n_e\approx 10^{-4}$ cm$^{-3}$ in the number density).
The parameter $\phi$ is determined after the simulation by fitting the radio flux of A399--A401 observed at $\approx$100 MHz, and it carries an uncertainty of approximately one order of magnitude (Sect.~\ref{sec:bridge_spect}).
Table~\ref{tab:params} summarizes the parameters in our fiducial model.
The dependence on the model parameters is discussed in Sect.~\ref{sec:discussion} and Sect.~\ref{sec:models}.
\par

\begin{figure}
    \centering
    \includegraphics[width=\linewidth]{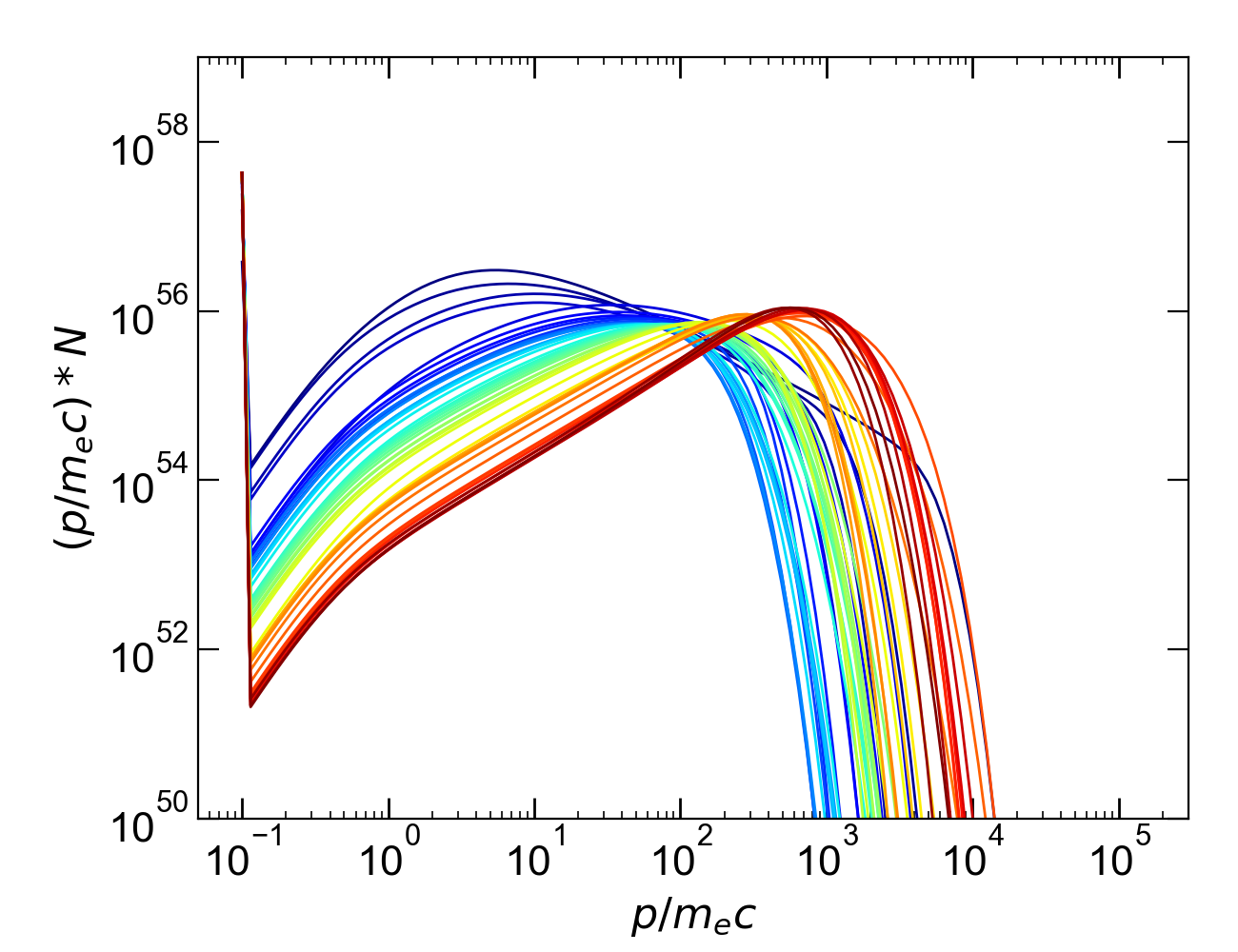}
    \caption{Example of CRE spectrum in a tracer simulated with our FP code. Shown is the evolution from $z =1$ to $z = 0.02$ (from blue to red).}
    \label{fig:CR_spect}
\end{figure}

At early times (blue curves), the spectra exhibit a low-energy cutoff at $p_{\rm min}^{\rm inj}/(m_{\rm e}c)=10$.

As an example, we show the evolution of the CRE spectrum (from blue to red) for a tracer particle in Fig.~\ref{fig:CR_spect}. We multiply $N(p)$ by momentum $p$ to show the number of CREs per logarithmic momentum interval. The tracer is randomly selected from the central region of the most massive cluster within the ROI. At early times (blue curves), the spectra are close to the initial condition and exhibit a low-energy cutoff at $p_{\rm min}^{\rm inj}/(m_{\rm e}c)=10$.
Low-energy CREs with $p/(m_{\rm e}c) < 10^2$ are strongly cooled by Coulomb interactions with ambient electrons, whereas the high-energy component with $p/(m_{\rm e}c) \gtrsim 10^3$ is dominated by radiative losses. 
According to the increase of the turbulent energy, seed CREs accumulated around $p/m_ec \sim 10^2$-10$^3$ are reaccelerated, forming a characteristic bump-shaped spectrum.

\begin{figure*}
    \centering
    
    \includegraphics[width=\linewidth]{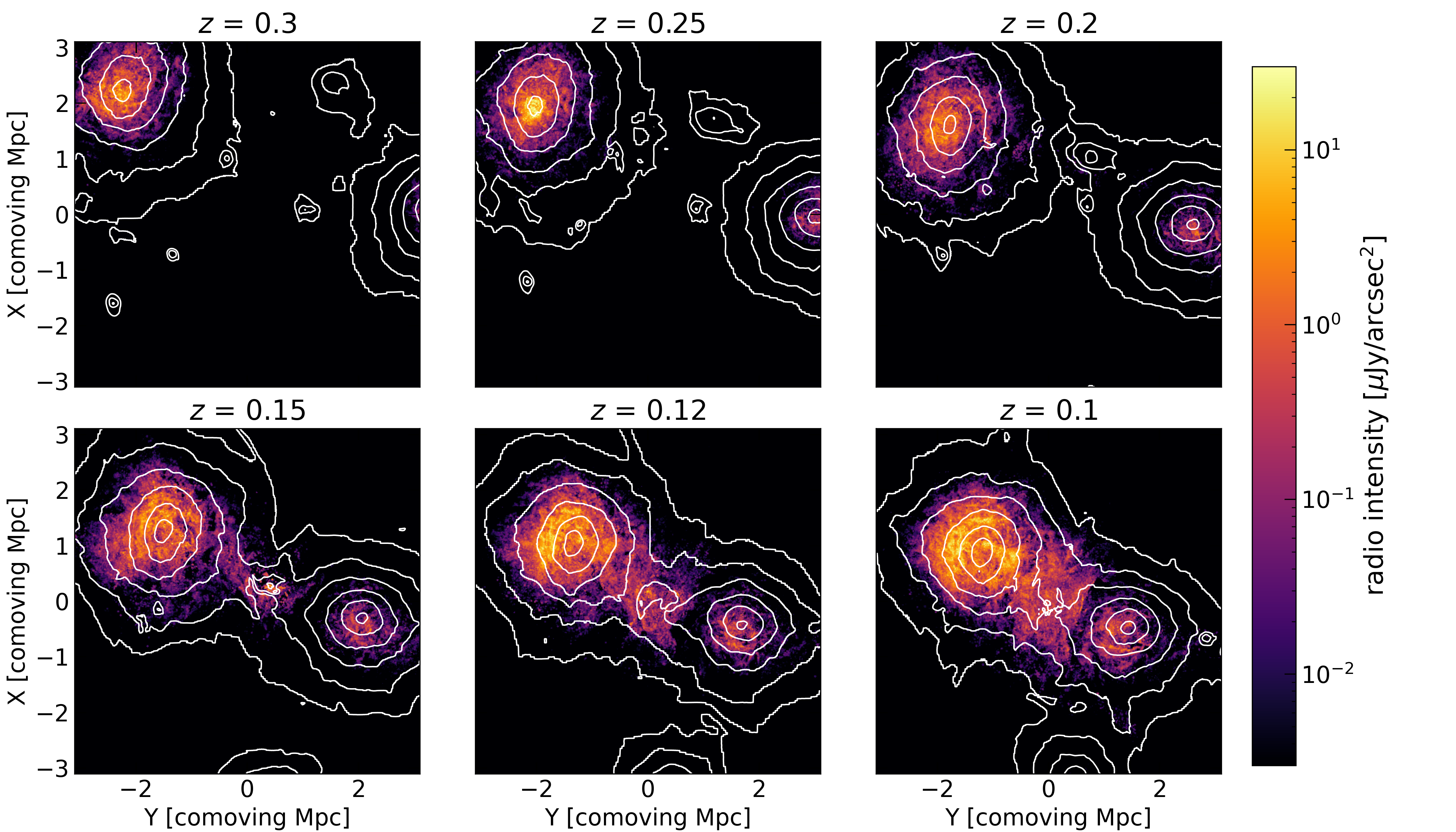}
    \caption{
    Evolution of the radio intensity map at 140 MHz. 
    The white contour shows the projected gas density at logarithmically spaced eight levels between $3\times10^{12}$ $M_\odot$/Mpc$^2$ and $2\times10^{14}$ $M_\odot$/Mpc$^2$. The mesh size of the radio map is 16 (comoving) kpc.
     }
    \label{fig:bridge_radio_map_z}
\end{figure*}

\begin{figure}
    \centering
    \includegraphics[width=\linewidth]{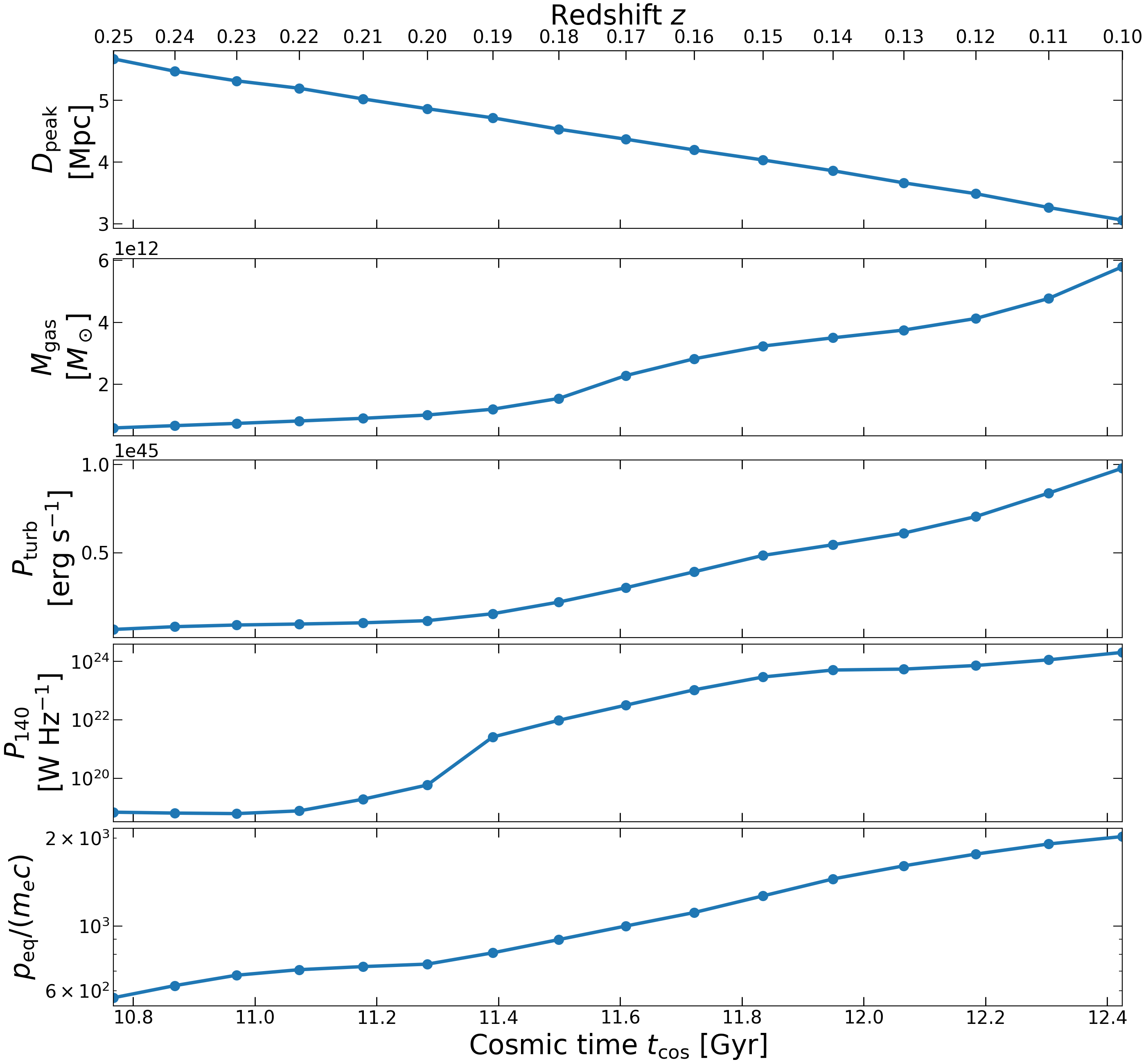}
    \caption{
    Evolution of bridge properties. From top to bottom, we show the time evolution of the X-ray peak distance between the clusters, baryon mass, solenoidal turbulent energy flux, and the 140 MHz radio power, and the typical equilibrium energy of CREs. The latter four quantities are calculated for the (1.2 Mpc)$^3$ box region extracted from the inter-cluster filament (Fig.~\ref{fig:Enzo_map_z}).
    }
    \label{fig:bridge_evolution}
\end{figure}

\section{Results \label{sec:results}}
In this section, we present the simulated synchrotron emission for our fiducial model. In Sect.~\ref{sec:bridge}, we study the evolution of turbulence in the bridge region and discuss the onset of the diffuse emission.
In Sect.~\ref{sec:bridge_spect}, we show that the spectrum and the spectral index distribution of the radio synchrotron emission are compatible with the observations of A399--A401.
The one-dimensional profiles of turbulence and radio emission in the bridge region are studied in Sect.~\ref{sec:bridge_profile}.
We also discuss the point--point correlation between radio and X-ray intensities in Sect.~\ref{sec:point-point}.

\subsection{Formation of a radio bridge \label{sec:bridge}}
Using the CR spectrum simulated with the FP solver, we calculate the synchrotron emission for each tracer particle.
To construct the emission map, we first prepare a three-dimensional mesh grid with a fixed resolution and sum the synchrotron emissivities of the tracers in each mesh cell. The intensity is then calculated by integrating the emissivity of each cell along the line of sight (LOS). Details of the tracer deposition method are given in Appendix~\ref{app:trc_deposit}.
\par

Fig.~\ref{fig:bridge_radio_map_z} shows the simulated synchrotron intensity map of the bridge region at 140 MHz.
We denote the 140 MHz intensity as $I_{140}$.
We use a mesh grid with a spatial resolution of 16 kpc and compute the projection along the Z axis of the original Enzo simulation.
The radio maps along the X axis are presented in Fig.~\ref{fig:bridge_radio_map_x}.
The merger axis is nearly aligned with the Y axis, so the separation between the two clusters is similar in both projections.
The white contour shows the projected gas density map, the same as in Fig.~\ref{fig:Enzo_map_z}.
The radio emission connecting two clusters starts to form at a redshift of $z\approx0.2$. At that epoch, a mass clump approaching from the (+X, +Y, $-$Z) direction of the simulation begins to collide with the filamentary gas bridging the cluster pair.
At $z = 0.15$, another very small clump collides with the filament in the -Z direction (see Fig.~\ref{fig:bridge_radio_map_x}).
The intensity of the bridge emission exhibits an increase following the collisions of those clumps.
Around $z = 0.12$, the bridge emission extends over an area of several square megaparsecs with a 140 MHz intensity $I_{140}$ exceeding the LOFAR sensitivity, i.e., $I_{140} \gtrsim 0.2$~$\mu$Jy~arcsec$^{-2}$ \citep{deJong_2022}.
Note that there is a radio halo forming in the primary cluster on the left side of each panel. This halo is initiated by a merger event occurring at $z\approx 0.5$. 
The radio intensity in the halo region is greater than the bridge intensity by a factor of $\approx$10 (Sect.~\ref{sec:bridge_profile}).
The evolution of the radio luminosity of the halo is discussed in Sect.~\ref{sec:halo}.
\par

To study the properties of the radio bridge, we extract a box-shaped region in the inter-cluster filament with a volume of $(1.2~{\rm Mpc})^3$.
The region is denoted by the white dashed square in Fig.~\ref{fig:Enzo_map_z}.
We show the time evolution of the integrated quantities in the box in Fig.~\ref{fig:bridge_evolution}, such as the mass of baryonic gas, the turbulent kinetic power, and the radio power.
The top row shows the projected distance between the two clusters measured from the peaks of X-ray intensity.
The merger of the clump causes a rapid increase in the mass in $0.2<z<0.15$. 
We find that the power of solenoidal turbulence also increases around this epoch.
The radio power at a fixed frequency, 140 MHz, is nonlinearly dependent on the turbulent power and evolves by more than four orders of magnitude from $z = 0.2$ to $z= 0.1$.
\par

In the bottom row of Fig.~\ref{fig:bridge_evolution}, we show the time evolution of the characteristic cut-off momentum, $p_{\rm eq}$, of the CRE spectrum. We define $p_{\rm eq}$ as the momentum at which the radiative cooling time equals the reacceleration timescale, $t_{\rm acc} = p^2/4D_{pp}$ (Eq.~\ref{eq:Dpp}), that is, $t_{\rm cool}(p_{\rm eq}) = t_{\rm acc}$. 
For this estimate, we use the mean values of gas density, temperature, and turbulent velocity measured within the box region to compute the relevant timescales. The evolution of the radio power is associated with the increase in $p_{\rm eq}$.
\par

After the turbulent power reaches $P_{\rm turb}\approx 5\times10^{44}$ erg~s$^{-1}$, the reacceleration becomes efficient and compensates for the radiative energy loss of CREs emitting $\approx$100 MHz radiation.
After $z \simeq 0.15$, when the projected separation decreases to $D_{\rm peak} \lesssim 4$~Mpc, the compression of the inter-cluster filament induced by the approaching motion between the two clusters becomes increasingly important for the turbulent power in the bridge region in addition to the clump accretion around $0.2<z<0.15$.

The radio power increases by more than two orders of magnitude in the last 1 Gyr. This rapid evolution reflects the timescale of turbulent reacceleration, which is $t_{\rm acc} \approx 500$ Myr. 
The 140 MHz power reaches $P_{140} \approx 4.5\times10^{24}$ W~Hz$^{-1}$ at $z\approx0.1$ when the separation of the two clusters becomes $\approx 3$ Mpc. This value is similar to the radio power of the A399--A401 radio bridge, $P_{140} \approx 1.0\times10^{25}$ W Hz$^{-1}$, as reported in \citet{Govoni_2019}.

\begin{figure*}
    \centering
    \begin{minipage}{0.48\linewidth}
        \includegraphics[width=\linewidth]{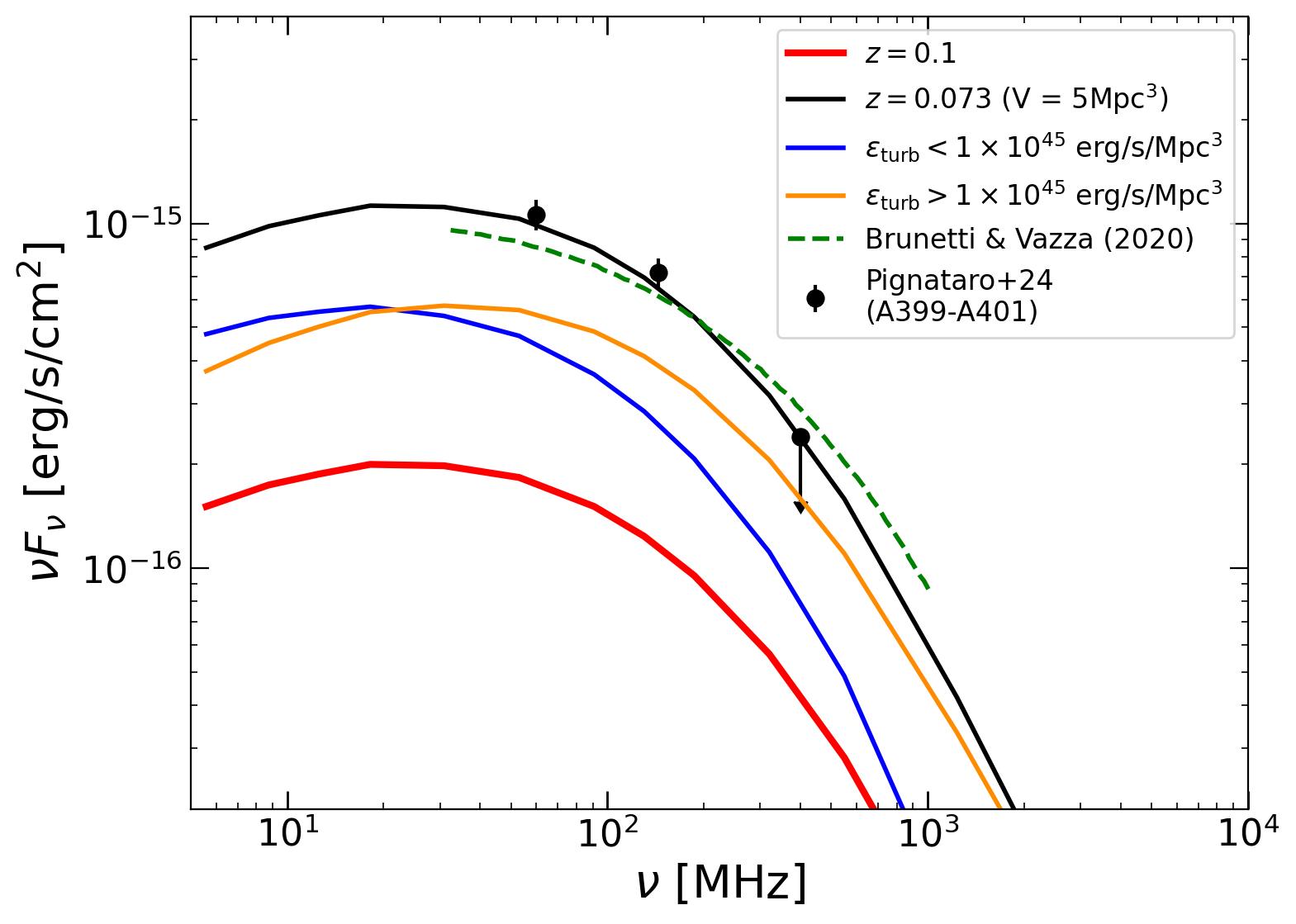}
    \end{minipage}
    \begin{minipage}{0.48\linewidth}
        \includegraphics[width=\linewidth]{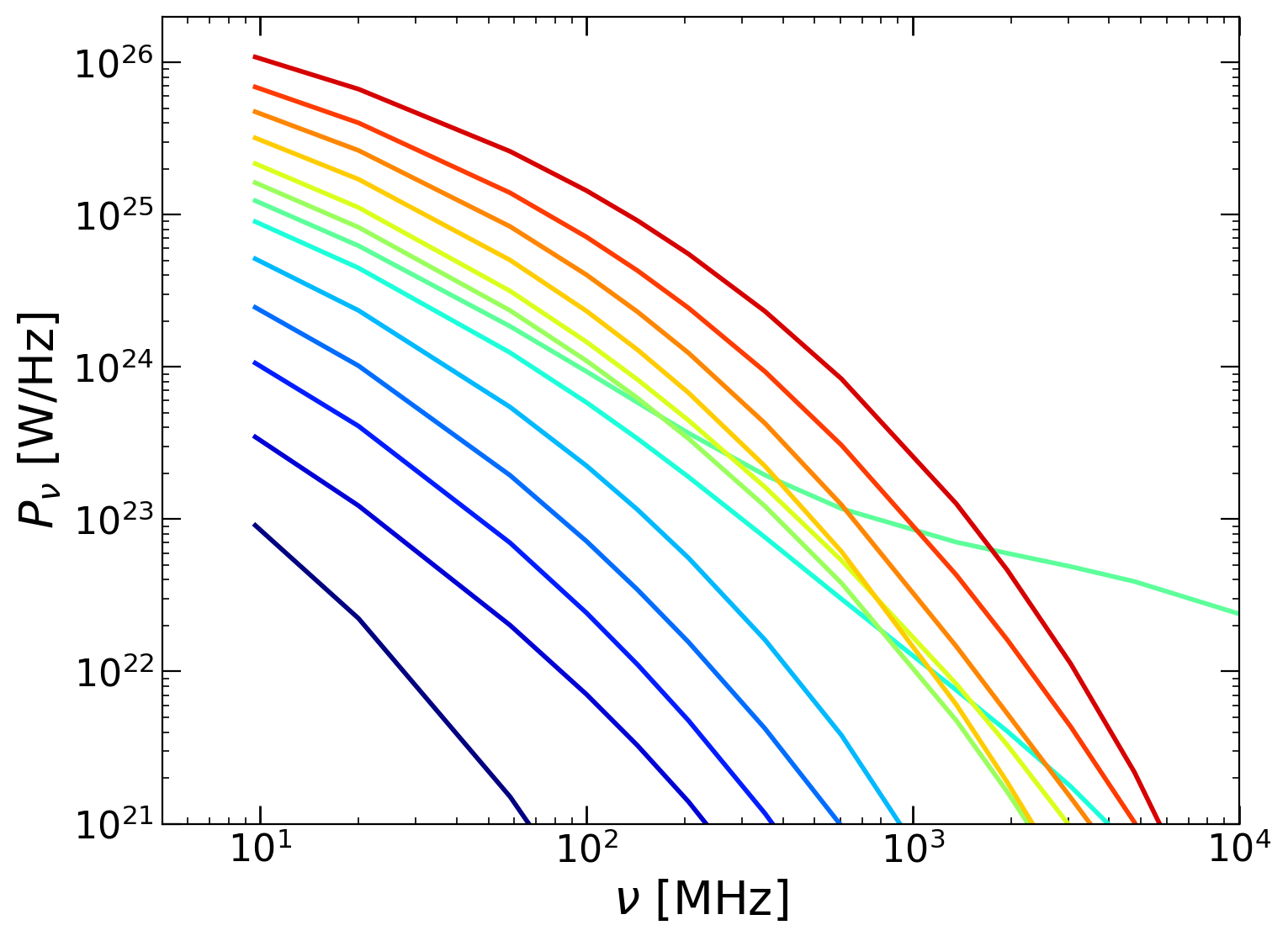}
    \end{minipage}
    \caption{
    Synchrotron spectrum of the radio bridge. In the left panel, we show the spectrum calculated for the snapshot at $z = 0.1$. The red and black lines show the flux calculated with the luminosity distances for $z=0.1$ and $z=0.073$, respectively. The data points with error bars show the flux of A399--A401 bridge measured at LOFAR frequencies, taken from \citet{Pignataro_2024}. The orange and blue lines show contribution from regions with volumetric turbulent energy flux larger and smaller than $1\times10^{45}$ erg/s/Mpc$^3$. The green dashed line shows the spectrum calculated by \citet{BV20}, adopted from \citet{Pignataro_2024_uGMRT}.
    The right panel shows the evolution of the emitted synchrotron power as a function of frequency. With the colors from blue to red, we plot the evolution from $z = 0.2$ to $z = 0.08$ (every $\Delta z = 0.01$). In the right panel, the spectra are shown in units of W Hz$^{-1}$ so that results at different redshifts (i.e., luminosity distance) can be directly compared.
    }
    \label{fig:bridge_spectrum}
\end{figure*}

\subsection{Synchrotron spectrum and spectral index \label{sec:bridge_spect}}

\subsubsection{Integrated spectrum}

Using the emission integrated in the extracted box, we study the spectrum of the radio bridge.
In the left panel of Fig.~\ref{fig:bridge_spectrum}, we show the spectrum calculated at the snapshot at $z = 0.1$ when the projected cluster separation is comparable to that of A399--A401. The one shown with the red line is calculated with the luminosity distance of $z=0.1$ ($D_{\rm L} \simeq 460$ Mpc) and the original volume of the box region, i.e., (1.2 Mpc)$^3 =$ 1.728 Mpc$^3$. 

We rescale the spectrum using the luminosity distance for the redshift of A399--A401 ($z\simeq 0.073$, $D_{\rm L} \simeq 330$ Mpc) and the approximate volume of the A399--A401 radio bridge \citep[$\approx$5 Mpc$^3$,][]{Govoni_2019, BV20}. 
The data points are LOFAR LBA (60 MHz) and HBA (144 MHz) fluxes of the A399--A401 bridge reported in \citet{Pignataro_2024}, which were derived by integrating the emission over an area of $\sim2.2$ Mpc$^2$. We note that the uncertainty in the LOS width of the bridge introduces an uncertainty in $\phi$. The observed spectral index between these two frequencies is $\alpha_{60}^{144} = -1.44\pm0.16$. The arrow shows the upper limit at 400 MHz given by \citet{Pignataro_2024_uGMRT}. 
The overall normalization of the simulated spectrum can be controlled with the parameter $\phi$ (Table~\ref{tab:params}). In the fiducial model, we choose $\phi = 6\times10^{-8}$ so that the rescaled 60 MHz intensity shown with the black dashed line matches the observed one.
Note that the value of $\phi$ depends on the choice of the snapshot used for normalization. As shown in the right panel of Fig.~\ref{fig:bridge_spectrum}, the radio power at around 100 MHz varies by about an order of magnitude in the redshift range $0.13 \leq z \leq 0.08$. The radio map looks similar to the observed radio bridge during this range. Thus, the reported value of $\phi$ should have an uncertainty of at least a factor of ten.
As discussed in Sect.~\ref{sec:models}, $\phi$ also depends on the model parameters, such as $\eta_B$ and $\psi$.
\par

We find that the radio bridge has a steep ($\alpha \approx -1.5$) spectrum at $\nu\approx100$ MHz and the spectral shape is similar to that found in the A399--A401 bridge.
For comparison, we also plot the spectrum calculated using the model of \citet{BV20} in Fig.~\ref{fig:bridge_spectrum} (green dashed line), adopted from \citet{Pignataro_2024_uGMRT}. They adopted the same parameter values, $\eta_B = 0.05$ and $\psi = 0.5$, as in our fiducial model. 
Unlike our tracer-based approach, their calculations are based on the turbulent conditions extracted from a single snapshot of the simulation presented in \citet{Govoni_2019}, and do not follow the long-term evolution of the dynamics or reacceleration and cooling of CR\footnote{The treatment of highly turbulent cells differs between the two approaches: we impose an upper limit on the turbulent Mach number (Sect.~\ref{sec:sim_turb}), whereas \citet{Pignataro_2024_uGMRT} adopted a lower limit on the reacceleration timescale, $t_{\rm acc} > 200$~Myr. 
Despite these methodological differences, the resulting spectra are broadly consistent.}.

\par

Following \citet{BV20}, we study the contribution from the regions with large and small turbulent power density (orange and blue dashed lines). We define the volumetric turbulent energy power $\epsilon_{\rm turb}$ in units of erg s$^{-1}$ Mpc$^{-3}$. 
We find that the emission above 100 MHz is dominated by cells with high $\epsilon_{\rm turb}$ ($>1\times10^{45}$ erg~s$^{-1}$~Mpc$^{-3}$), which occupy only $\approx 18\%$ of the volume. This result is in qualitative agreement with \citet{BV20}.
The dominance of the high-$\epsilon_{\rm turb}$ regions is more pronounced at higher frequencies, which implies that the radio image becomes less smooth at higher frequencies.
\par

In the right panel of Fig.~\ref{fig:bridge_spectrum}, we show the time evolution of the radio spectrum from $z = 0.2$ to $z = 0.08$. 
The radio power around 100 MHz evolves very rapidly, and the spectral shape shows gradual flattening.
At $z = 0.15$ and $z = 0.14$ (light-green lines), the spectrum above 200 MHz suddenly becomes very flat.
The impact of the mass clump induces intense turbulence around it, which leads to efficient reacceleration. The enhancement of the emission around the clump can also be seen in the radio map (Fig.~\ref{fig:bridge_radio_map_z}). This flattening feature is evanescent and decays in $200$ Myr.
Note that the reacceleration efficiency in that region can be affected by the upper limit that we impose on the turbulent Mach number (Sect.~\ref{sec:sim_turb}).
\par

\begin{figure*}
    \centering
    \begin{minipage}{0.48\linewidth}
        \includegraphics[width=\linewidth]{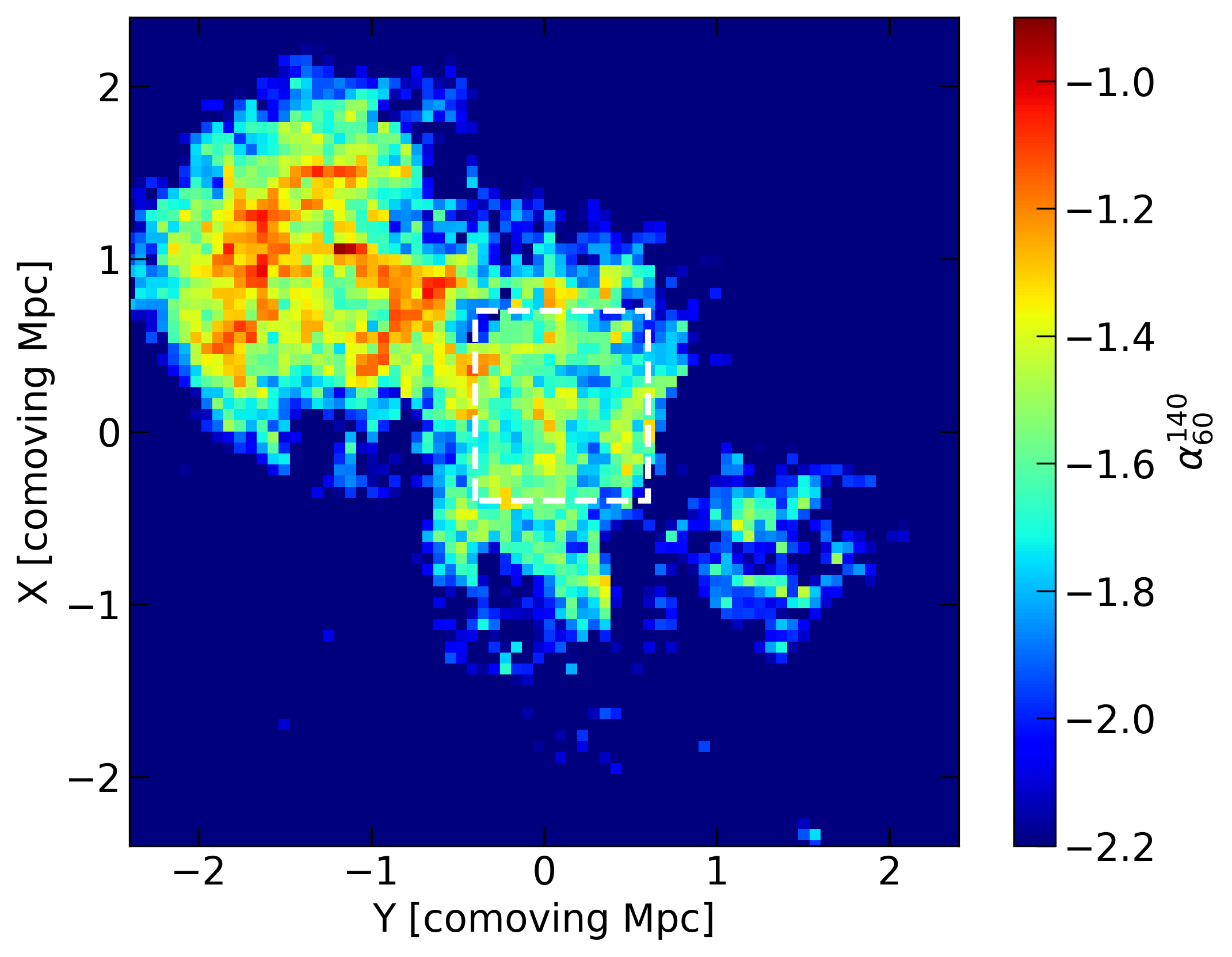}
    \end{minipage}
    \begin{minipage}{0.48\linewidth}
        \includegraphics[width=\linewidth]{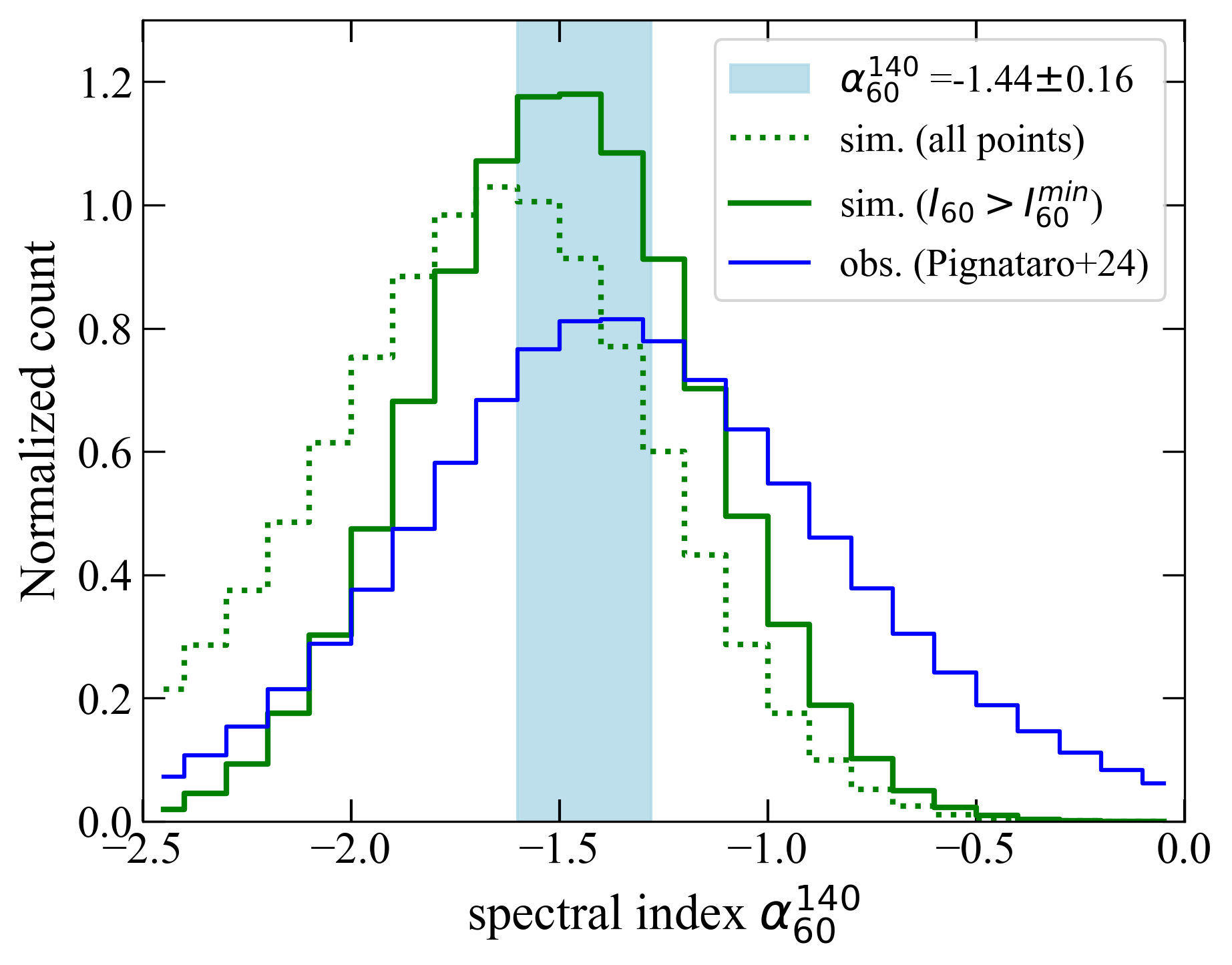}
    \end{minipage}
    \caption{
    Synchrotron spectral index between 60 MHz and 140 MHz. Left panel: spectral index map at a spatial resolution of $64\,{\rm kpc} \times 64\,{\rm kpc}$ over a $(4.8~{\rm comoving~Mpc})^2$ region centered on the radio bridge. The white box indicates the bridge region, identical to that shown in Fig.~\ref{fig:Enzo_map_z}. Right panel: spectral index distribution within the bridge region. The green histograms show the simulation results, where the dotted line corresponds to the unweighted distribution and the solid line to the distribution weighted by the 60 MHz intensity. The blue histogram represents the observed distribution reported by \citet{Pignataro_2024}. The shaded blue region indicates the spectral index derived from the integrated flux in the observation. All histograms are normalized to unit area.
    }
    \label{fig:spix}
\end{figure*}

\subsubsection{Spectral index map}\label{sec:spix}
Observations of the spectral index distribution of the radio emission provide important constraints on the formation scenario of diffuse radio bridges.
Concerning the radio bridge of A399--A401, \citet{Pignataro_2024} compared the LOFAR Low Band Antenna (LBA) data at 60 MHz with the High Band Antenna (HBA) data at 140 MHz and found that the spectral index $\alpha_{60}^{140}$ is as steep as  $\alpha_{60}^{140}\approx -1.4$, with no significant spatial trend across or along the bridge.
\par

In the left panel of Fig.~\ref{fig:spix}, we present the simulated map of the spectral index $\alpha_{60}^{140}$ measured between 60 MHz and 140 MHz in the bridge region. The map shows a $(4.8~{\rm comoving~Mpc})^2$ region centered on the radio bridge.
The white square denotes the region of the radio bridge, which is the same as the one shown in Fig.~\ref{fig:Enzo_map_z}.
The right panel presents the distribution of the spectral index within this region.
The dotted and solid green lines correspond to the spectral index distributions for all pixels in the region and for those above the LOFAR sensitivity threshold at 60 MHz, respectively.
We use $I_{60}^{\rm min} = 1.2\;\mu$Jy~arcsec$^{-2}$ as the threshold intensity at 60 MHz \citep[][]{Pignataro_2024}.
The histograms are plotted with a bin width of $\Delta\alpha_{60}^{140} = 0.1$.
We use the data adopted from Fig.~4 of \citet{Pignataro_2024} to plot the observational histogram shown in blue.
Since the bin width of the histogram is smaller than the error in the observed $\alpha_{60}^{140}$, which is typically 0.3, we distribute each observed $\alpha_{60}^{140}$ over the histogram using a Gaussian kernel whose width corresponds to its observational error $\Delta\alpha_{60}^{140}$.
We also smooth the distribution of the simulated spectral index in the same way, but using a fixed kernel width of $\Delta\alpha_{60}^{140}=0.3$.
We normalize each histogram such that its integrated area becomes unity to compare the shapes of the distributions.
\par

We find that both the observed and simulated distributions have peaks around $\alpha_{60}^{140} \approx -1.5$. 
The simulated distribution is shifted to slightly larger $\alpha_{60}^{140}$ when the observational sensitivity is taken into account.
The observed distribution is more extended toward the large-$\alpha_{60}^{140}$ (flat-spectrum) than the simulated one\footnote{Note that the observational error is relatively large for the flat-spectrum points ($\alpha_{60}^{140} > -1$) compared to the steeper spectrum ones. See Fig.~2 of \citet{Pignataro_2024}.}.
One of the possible causes for this lack of flat spectra would be our simplified assumption on the CR injection.
Although we only consider an aged population of CREs injected at $z = 1$, there should also be CREs freshly accelerated at shocks and AGNs at later times. 
The mixing of continuous injection and reacceleration is expected to generate different components with a broader spectral distribution and a stretched spectrum \citep[e.g.,][]{Brunetti_2017,Nishiwaki_2021}.
In addition, our magnetic-field dynamo model may underestimate the dispersion of the magnetic field strength in the bridge region compared to the resolved dynamo in the MHD simulation (see Fig.~\ref{fig:B_tacc_evo}). 
A larger field fluctuation may lead to a broader distribution of the spectral index.
\par

\begin{figure*}
    \centering
      \begin{minipage}{0.48\linewidth}
    \centering
    \includegraphics[width=\linewidth]{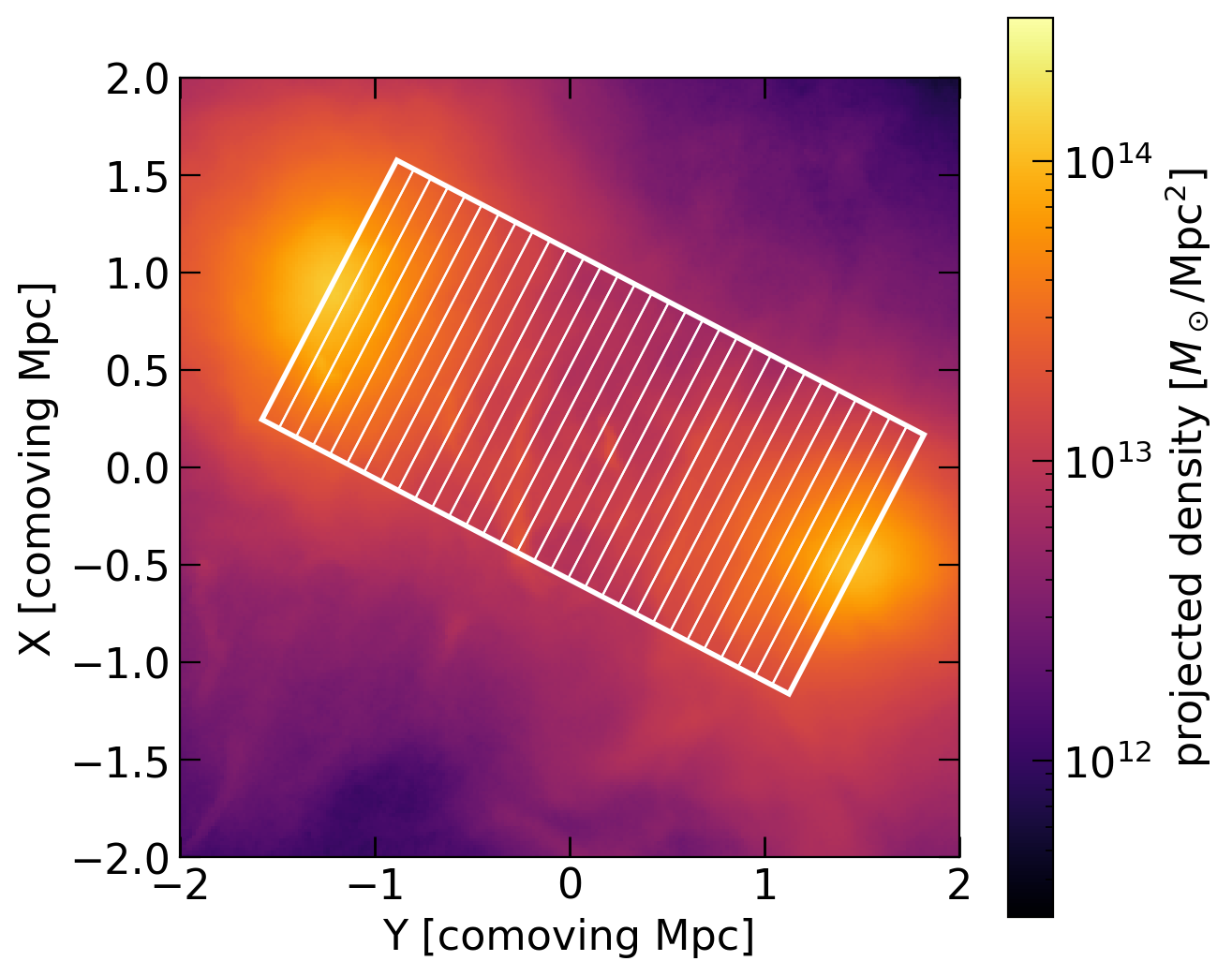}
  \end{minipage}
   \begin{minipage}{0.48\linewidth}
    \centering
    \includegraphics[width=\columnwidth]{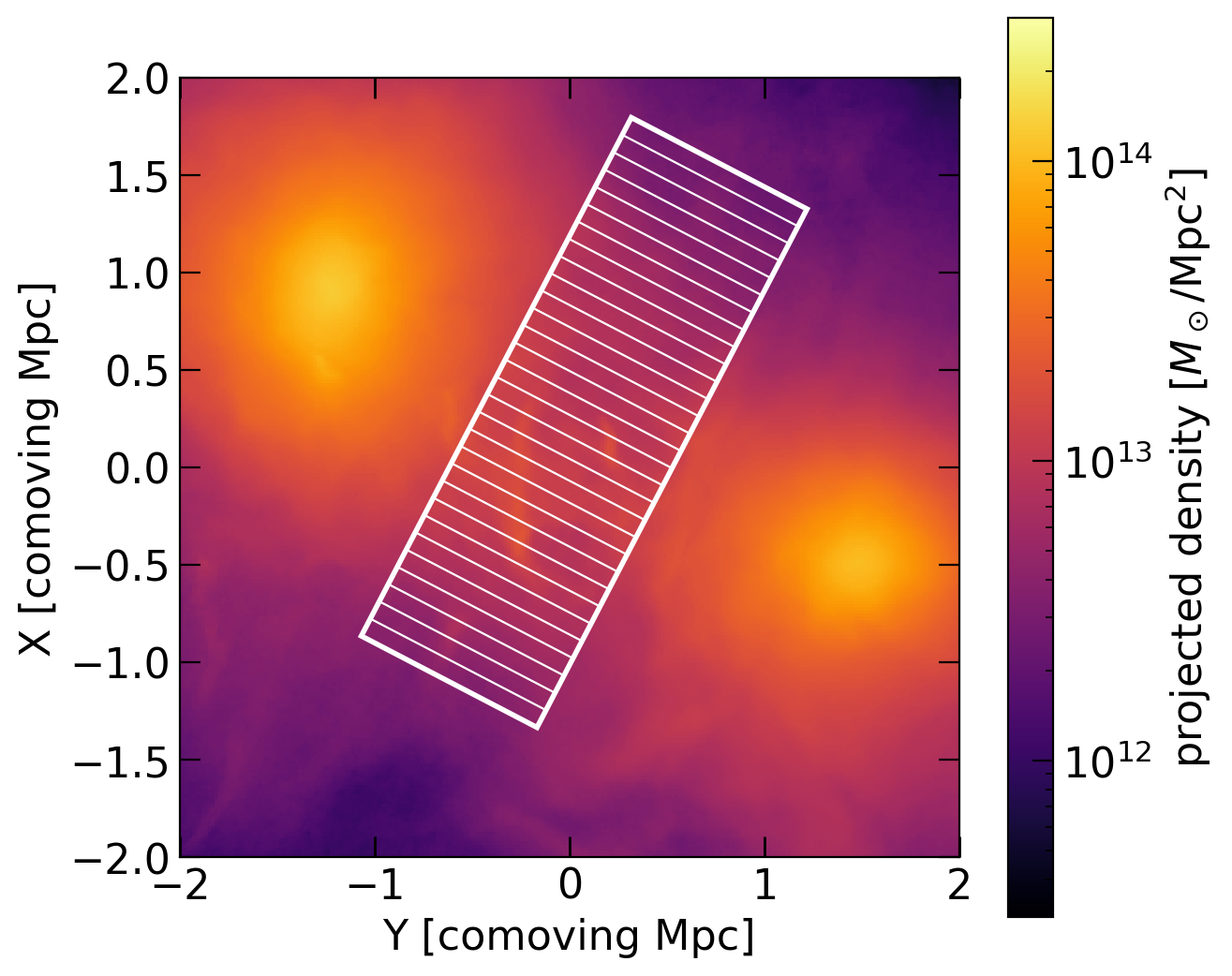}
  \end{minipage}
  \begin{minipage}{0.48\linewidth}
    \centering
    \includegraphics[width=\columnwidth]{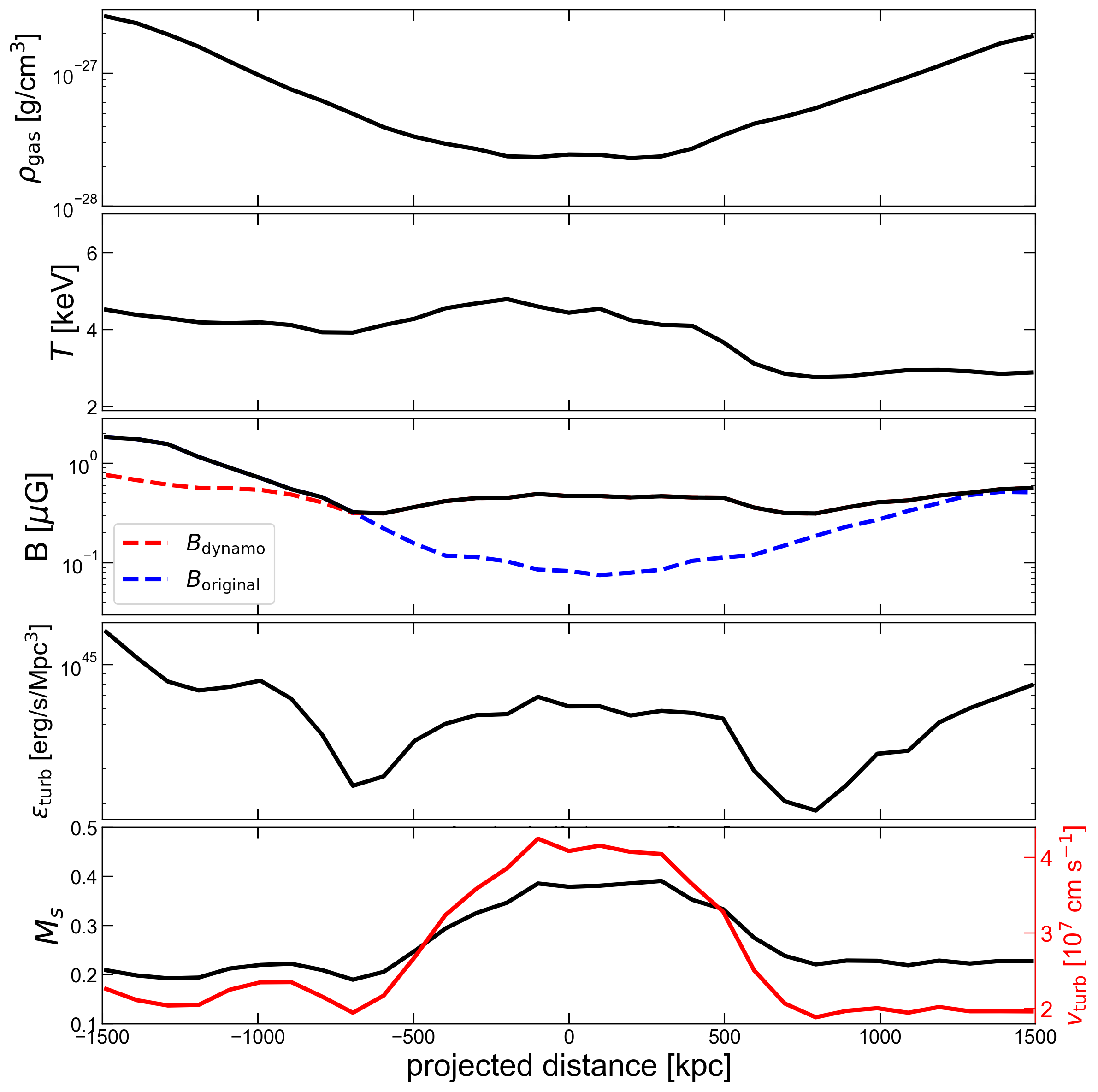}
  \end{minipage}
  \begin{minipage}{0.48\linewidth}
    \centering
    \includegraphics[width=\columnwidth]{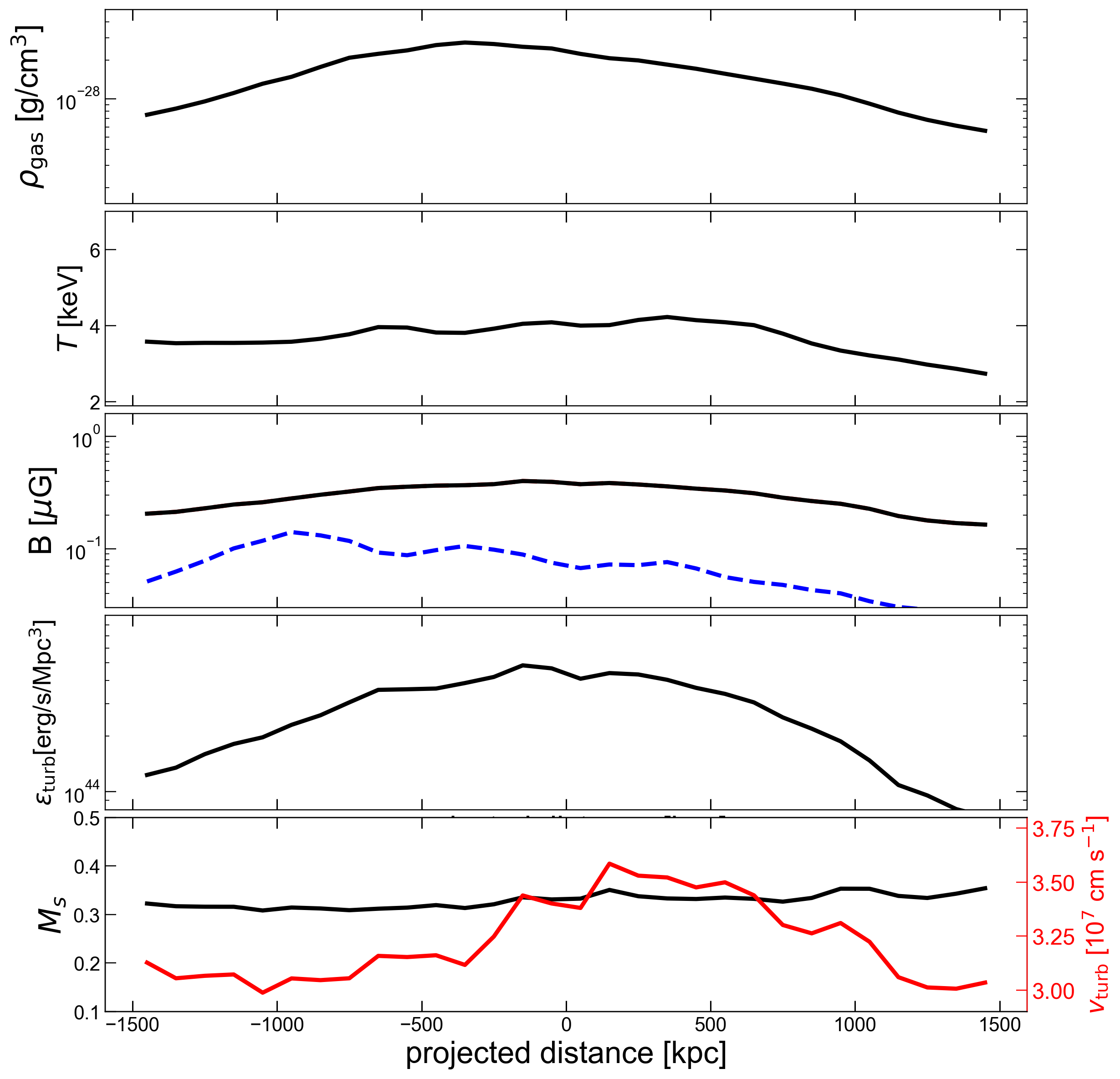}
  \end{minipage}
  \begin{minipage}{0.48\linewidth}
    \centering
    \includegraphics[width=\columnwidth]{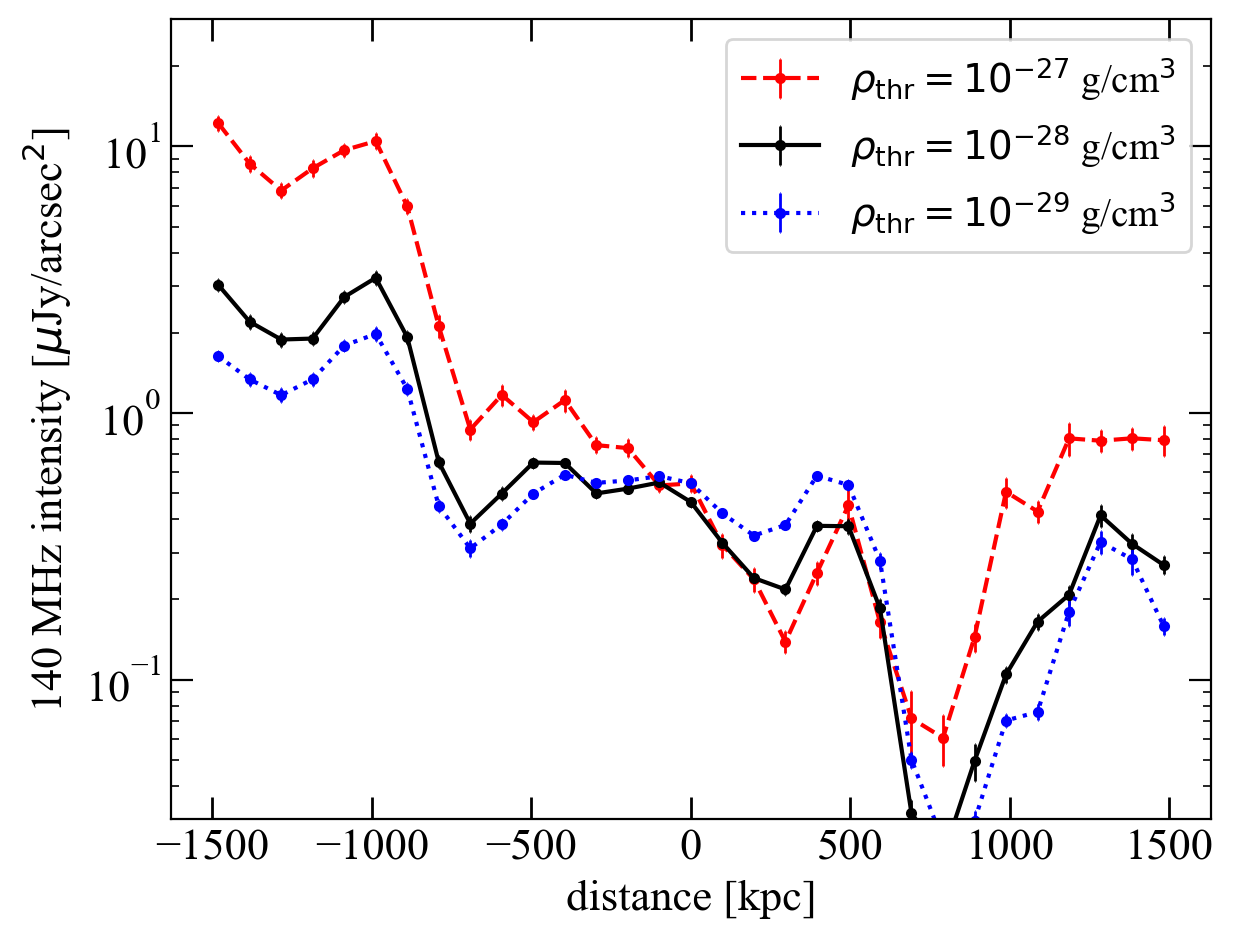}
  \end{minipage}
  \begin{minipage}{0.48\linewidth}
    \centering
    \includegraphics[width=\columnwidth]{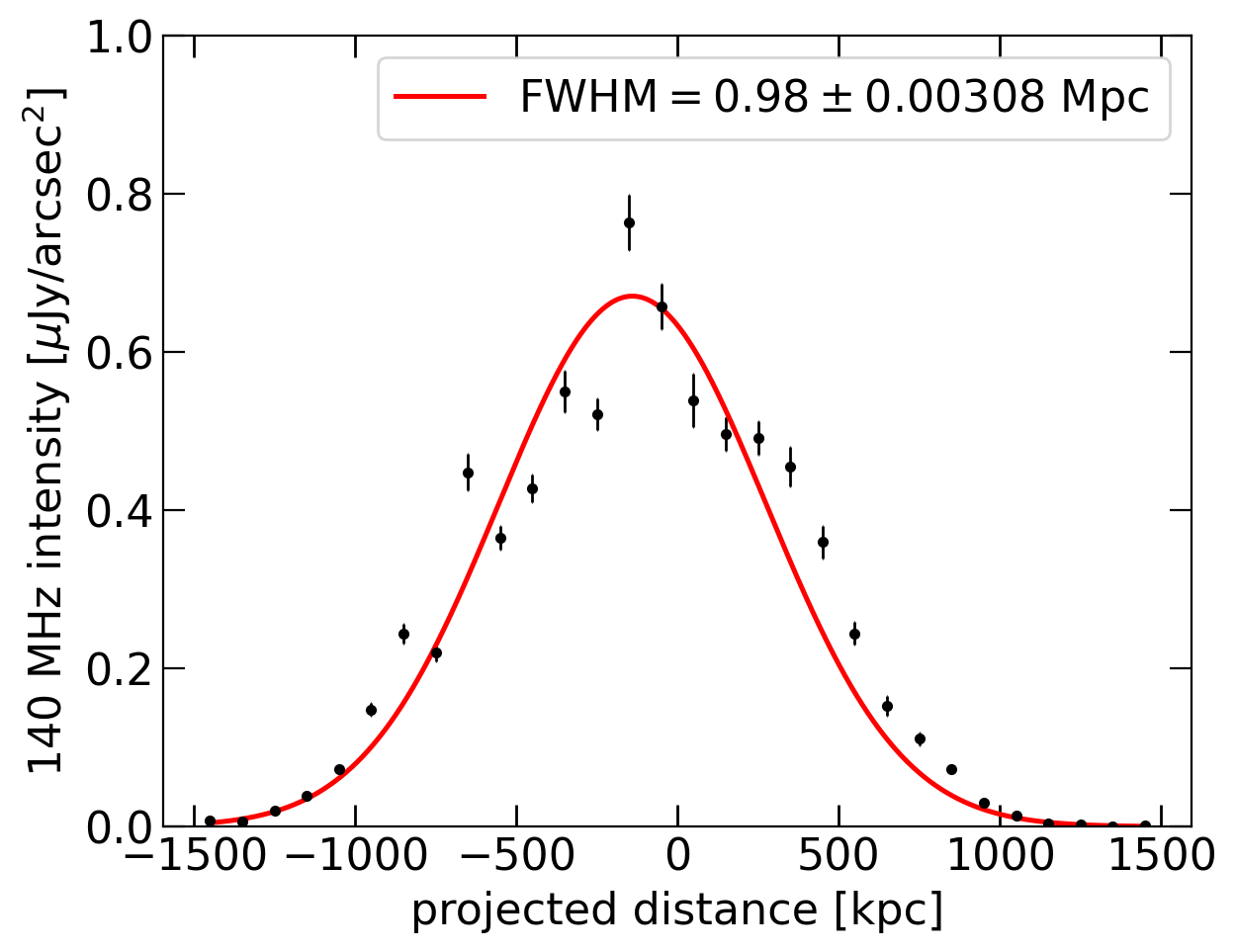}
  \end{minipage}
    \caption{
    Profiles of the radio bridge parallel to (left panels) and perpendicular to (right panels) the merger axis.
    The top row shows the projected gas density map, where the white squares and stripes indicate the regions used to compute the profiles.
    The second row presents the profiles of gas density, temperature, magnetic field, solenoidal turbulent power per unit volume, and turbulent sonic Mach number together with the turbulent velocity.
    For the magnetic field, the red dashed line shows the strength estimated with the dynamo model with $(L,\eta_B) = (150~{\rm kpc},0.05)$, while the blue dashed line shows the magnetic field taken from the original Enzo snapshot.
    The solid black line indicates the larger value between the two magnetic fields, which is adopted in the FP simulation (Sect.~\ref{sec:dynamo}).
    In the Mach number panel, the red solid line shows the solenoidal turbulent velocity $v_{\rm turb}$ (right axis).
    The third row shows the projected radio intensity profile.
    The error bars represent the $1\sigma$ uncertainty of the mean value.
    }
    \label{fig:bridge_profiles}
\end{figure*}

\subsection{Bridge profiles \label{sec:bridge_profile}}
In this section, we investigate one-dimensional profiles in the region of the radio bridge. We consider the profiles in two directions, parallel to and perpendicular to the line connecting to the projected centers of the two clusters. For simplicity, we refer to that line as the ``merger axis.''
As in the previous sections, we consider the snapshot at $z = 0.1$ and its projection along the Z axis.
\par

To study the profiles parallel to the merger axis, we consider a rectangular region that has a length equal to the projected distance between the clusters and a width of 1.5 Mpc (the top left panel of Fig.~\ref{fig:bridge_profiles}). The rectangular region is divided into contiguous stripes with a fixed width of 100 kpc along the merger axis.
We consider a depth of 1.5 Mpc along the LOS for each stripe and calculate the profiles of gas density, temperature, the magnetic field, and the volumetric power and sonic Mach number of the solenoidal turbulence, using quantities averaged within the volume of each stripe.
The profiles are shown in the middle panels of Fig.~\ref{fig:bridge_profiles}.
We find that the bridge region has a gas density of $\rho_{\rm gas} \approx 2\times10^{-28}$ g~cm$^{-3}$, which corresponds to a thermal electron density of $n_e\approx2\times10^{-4}$ cm$^{-3}$ when the mean molecular weight is $\mu_{\rm mol} \approx 0.6$.
The typical temperature is 4 keV, which is comparable to but slightly smaller than the temperature of the bridge region of A399--A401 measured from X-ray observations \citep[6.5 keV,][]{Akamatsu_2017}.
The typical magnetic field strength is $B_{\rm dyn}\approx0.5$~$\mu$G, which is larger than the field strength of the original Enzo data by a factor of $\approx3$ (see also Fig.~\ref{fig:B_tacc_evo}). The dotted line in the panel for $B_{\rm dyn}$ shows a profile proportional to $n_e^{0.5}$, which is often used to fit the magnetic field profile in ICM measured with the Faraday rotation measure \citep[e.g.,][]{Bonafede_2010}.
The magnetic field in the bridge region clearly exceeds that profile extrapolated from the cluster center. 
The turbulent power $\varepsilon_{\rm turb} = 1/2\rho_{\rm gas}\delta v^3/L$ in the bridge region is about half of that in the cluster center, although the gas density in the bridge is about ten times smaller than that in the cluster center.
The turbulent velocity averaged in each stripe region is about two times larger in the bridge region than the cluster region.
The Mach number is typically $M_{\rm s} \approx 0.2$ in the clusters, and it increases in the bridge to $M_{\rm s} \approx 0.4$.
Since the reacceleration efficiency strongly depends on the Mach number (Eq.~\ref{eq:Dpp}), CREs are efficiently reaccelerated in this region.
\par

We study the one-dimensional radio intensity profile by averaging the projected intensity over the area of each strip.
The radio intensity profile along the axis is shown in the bottom left panel of Fig.~\ref{fig:bridge_profiles}. The colors distinguish the models with different initial conditions for the CR seeding. The comparison between the models is discussed in Sect.~\ref{sec:models}. In our fiducial model (black points), the bridge region is characterized by a low radio intensity of $I_{140} \approx 0.5$~$\mu$Jy~arcsec$^{-2}$ with small fluctuation in the profile.
The brightness in the central region of the primary cluster is almost an order of magnitude larger than that of the bridge region.
There is an enhancement of the radio brightness around $\approx 500$ kpc from the cluster center, which is coincident with the increase of $\varepsilon_{\rm turb}$.
The amplification of the turbulence in similar regions has been reported in previous studies \citep[e.g.,][]{Vazza_2017,Wittor_2017}.
The approach of the two clusters causes converging flows and weak shocks, which amplify the vorticity.
In reality, prompt injection of CREs from AGNs can also affect the brightness profile, which is not taken into account in this simulation.
\par

The right panels of Fig.~\ref{fig:bridge_profiles} show the profiles perpendicular to the projected merger axis.
We consider a region with a width of 2 Mpc centered on the merger axis and a length of 1.2 Mpc parallel to the merger axis.
We calculate the mean values in strips spaced every 100 kpc.
Although the gas density and the magnetic field are peaked around the merger axis, the temperature and the turbulent Mach number are nearly constant.
The 140 MHz intensity also exhibits a peaked profile, although it shows scatter around the peak. Fitting the profile with a Gaussian function, we obtain the best-fit full width at half maximum (FWHM) of 0.98 Mpc.
The bridge width depends on the projected direction. When the system is projected along the X axis (Fig.~\ref{fig:bridge_radio_map_x}), the best-fit FWHM becomes 0.84 Mpc.

\begin{figure*}
    \centering
    \begin{minipage}{0.48\linewidth}
    \includegraphics[width=\linewidth]{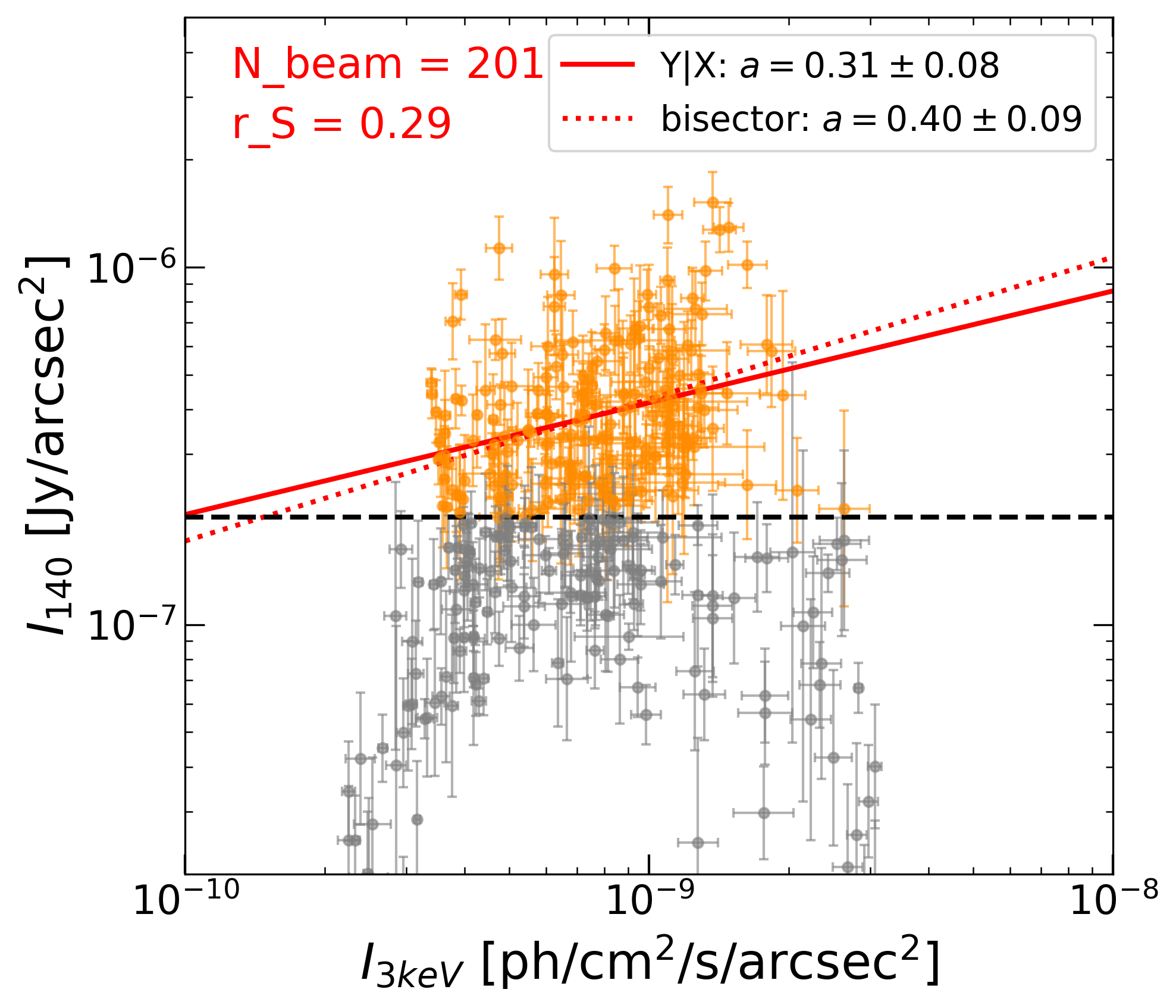}
    \end{minipage}
    \begin{minipage}{0.48\linewidth}
    \includegraphics[width=\linewidth]{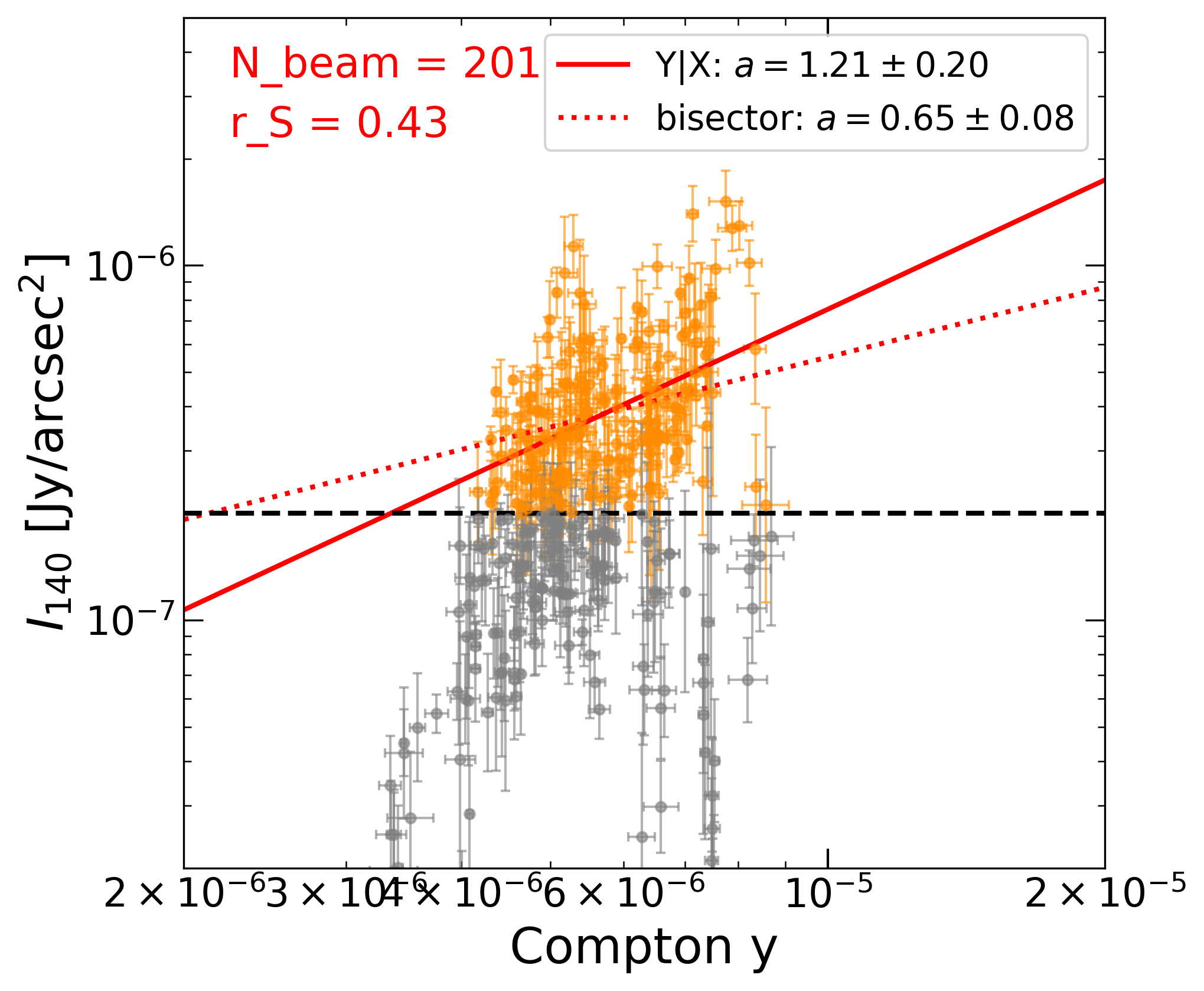}
    \end{minipage}
    \caption{
    Correlation plots for the radio and X-ray intensities (left panel) and for the radio intensity and the Compton-y parameter (right panel).
    We study the (1.2 Mpc)$^2$ area shown in Fig.~\ref{fig:Enzo_map_z}.
    The data points with error bars show the mean values and the standard deviation in the resampled pixels with 64 kpc resolution. The orange points have the 140 MHz intensity greater than the current sensitivity of LOFAR, $I_{140} \geq 0.2$ $\mu$Jy/arcsec$^2$ (dashed line).
    The red lines show the best-fit correlations for the orange points calculated with the BCES Y|X (solid line) and bisector (dotted line) methods.
    }
    \label{fig:point_point}
\end{figure*}

\subsection{Point-point correlations \label{sec:point-point}}

The pixel-by-pixel comparison of the radio and X-ray intensity maps provides important hints about the connection between thermal and nonthermal components in the ICM. 
Some classical radio halos in GCs are known to show sublinear correlations, i.e., $I_{\rm radio} \propto I_X^{a_{\rm X}}$ with $a_{\rm X} < 1$ \citep[e.g.,][]{Govoni_2001,Feretti_2001,Bonafede_2021}.
On the other hand, the correlation for mini halos tends to be super-linear ($a_{\rm X} > 1$) \citep[e.g.,][]{Ignesti_2020,Biava_2021}.
There are some exceptions, such as giant radio halos with super-linear correlations \citep[e.g.,][]{Rajpurohit_2018}.
The correlation in the radio bridge is still under debate due to the limited number of observations and data points available for each system.
Analyzing the LOFAR data of the A399--A401 bridge, \citet{deJong_2022} reported a weak sublinear correlation between the radio and X-ray intensities with a slope of $a_{\rm X} = 0.27\pm0.07$ and a Spearman correlation coefficient of $r_{\rm S} = 0.41\pm0.14$. 
A similar correlation is also found for the bridge in A1758 \citep[$a_{\rm X} = 0.25$, $r_{\rm S} = 0.52$,][]{Botteon_2020}, which is also considered a pre-merger system.
The point-to-point correlation between the radio intensity and the Compton-y parameter, which characterizes the thermal Sunyaev-Zeldovich (SZ) effect of ICM, has also been studied for a few nearby systems including A399--A401 \citep[e.g.,][]{Planck_2013_Coma,Radiconi_2022}.
Hereafter, we call the correlation between the radio and X-ray intensities and that between the radio intensity and the Compton-y parameter as the radio--X correlation and the radio--SZ correlation, respectively. 
\par

In this section, we explore the radio-X and radio-SZ correlations in the simulated radio bridge, using 140 MHz and 3 keV as the reference frequency and photon energy for the radio and X-ray emissions, respectively. 
For the X-ray emission, we ignore the line emission and only consider free-free emission for simplicity.
We calculate the radio and X-ray intensities on a mesh with a resolution of 16 kpc, as in Fig.~\ref{fig:bridge_radio_map_z}. 
The Compton-$y$ parameter is calculated as: 
\begin{equation}
    y = \int dl \frac{k_{\rm B}\sigma_T}{m_ec^2} n_{e}T,
\end{equation}
where $l$ is the distance along the LOS, $k_{\rm B}$ is the Boltzmann constant, and $\sigma_{\rm T}$ is the Thomson cross section.
We resample the data using 4$\times$4 neighboring pixels, producing a map with 64 kpc resolution.
As in the previous sections, we consider the snapshot at $z=0.1$ and the projection along the Z axis.
\par

In Fig.~\ref{fig:point_point}, we show the radio--X (left panel) and radio--SZ (right panel) point-to-point correlations.
As in the previous sections, we consider an area of (1.2 Mpc)$^2$ shown with the white box in Fig.~\ref{fig:Enzo_map_z}.
The points with error bars show the intensities or the Compton-y parameter averaged in the resampled pixels and their uncertainties.
Using the Bivariate Correlated Errors and intrinsic Scatter (BCES) method \citep[][]{Akritas_Bershady_1996}, we fit the correlation in log--log space with the power-law functions $\log(I_{140}) = a_{\rm X}\log(I_{\rm 3keV}) + b_{\rm X}$ and $\log(I_{140}) = a_{\rm Y}\log y + b_{\rm Y}$ for the radio--X and the radio--SZ correlations, respectively. 
We exclude the points with 140 MHz intensity below $I_{140}^{\rm min} = 0.2$~$\mu$Jy~arcsec$^{-2}$ (black dashed line) from the fitting, considering the sensitivity of the LOFAR observation \citep[e.g.,][]{deJong_2022}.
After this cut, $N_{\rm beam} = 201$ data points are used to study the correlations.

The fitting results for the BCES Y|X and BCES bisector methods are shown with the red solid line and red dotted lines, respectively. 
We find sublinear slopes for the radio--X correlation; $a_{\rm X}=0.31\pm0.08$ for BCES Y$|$X and $a_{\rm X}=0.40\pm0.09$ for BCES bisector. The Spearman correlation coefficient, $r_{\rm S} = 0.29$, suggests that the correlation is weak. The slopes are similar to the values reported for A399--A401 ($a_{\rm X}=0.27\pm0.07$) and A1758 ($a_{\rm X}=0.25\pm0.08$) \citep[][]{deJong_2022}.
The slope of the radio--SZ correlation significantly depends on the fitting method due to the narrow range of the Compton-y parameter in the bridge region.
The slope is super-linear when fitted with the BCES Y$|$X method ($a_{\rm Y}=1.21\pm0.20$), while it is sublinear with the BCES bisector method ($a_{\rm Y}=0.65\pm0.08$). The latter is close to the value found in the observation of the A399--A401 bridge, $a_{\rm Y}=0.62^{+0.16}_{-0.17}$, \citep[][]{Radiconi_2022}.
\par

In observations, the slope for the radio--X correlation in radio bridges is found to be smaller (flatter) than those reported for the radio halos. However, as discussed in \citet{Radiconi_2022}, the difference of the slope may not be intrinsic but may instead be caused by observational bias because of the limited sensitivities of observational facilities.
To test the effect of the sensitivity limit on the radio--X correlation, we perform the fitting using all data points (360 points) in Fig.~\ref{fig:point_point} including the gray ones below the sensitivity limit.
We find $a_{\rm X}=0.25\pm0.13$ for the BCES Y$|$X and $a_{\rm X}=0.18\pm0.09$ for the BCES bisector methods.
However, the fitting is seemingly affected by outliers with small $I_{140}$ and large $I_{\rm 3keV}$, which are coincident with the peripheral regions of two clusters intersecting with the extracted (1.2 Mpc)$^2$ area (see Fig.~\ref{fig:spix}).
When we exclude those outliers by limiting the area as (0.8 Mpc)$^2$ (with the same central position) and repeat the fitting, we get steeper slopes: $a_{\rm X}=0.67\pm0.15$ (BCES Y$|$X) and $a_{\rm X}=0.46\pm0.08$ (BCES bisector).
Notably, the slopes for all points in the (0.8 Mpc)$^2$ area are similar to those reported for the giant radio halos \citep[e.g.,][]{Govoni_2001,Duchesne_2021,deJong_2022}. 
Further observational and theoretical work is required to understand the intrinsic radio--X correlation in the radio bridge and the effect of limited sensitivity on the observed correlation.
\par

As a summary, our simulations based on the reacceleration model show results that are compatible with the observed radio--X correlation for radio bridges. 
The radio--SZ correlation is subject to large uncertainty due to the scatter in the radio intensity and the narrow range of the Compton-y parameter. 
The observed correlations can be heavily affected by the sensitivity of the observation and the selection of the ROI.

\section{Discussion \label{sec:discussion}}

\begin{figure}
    \centering
    \includegraphics[width=\linewidth]{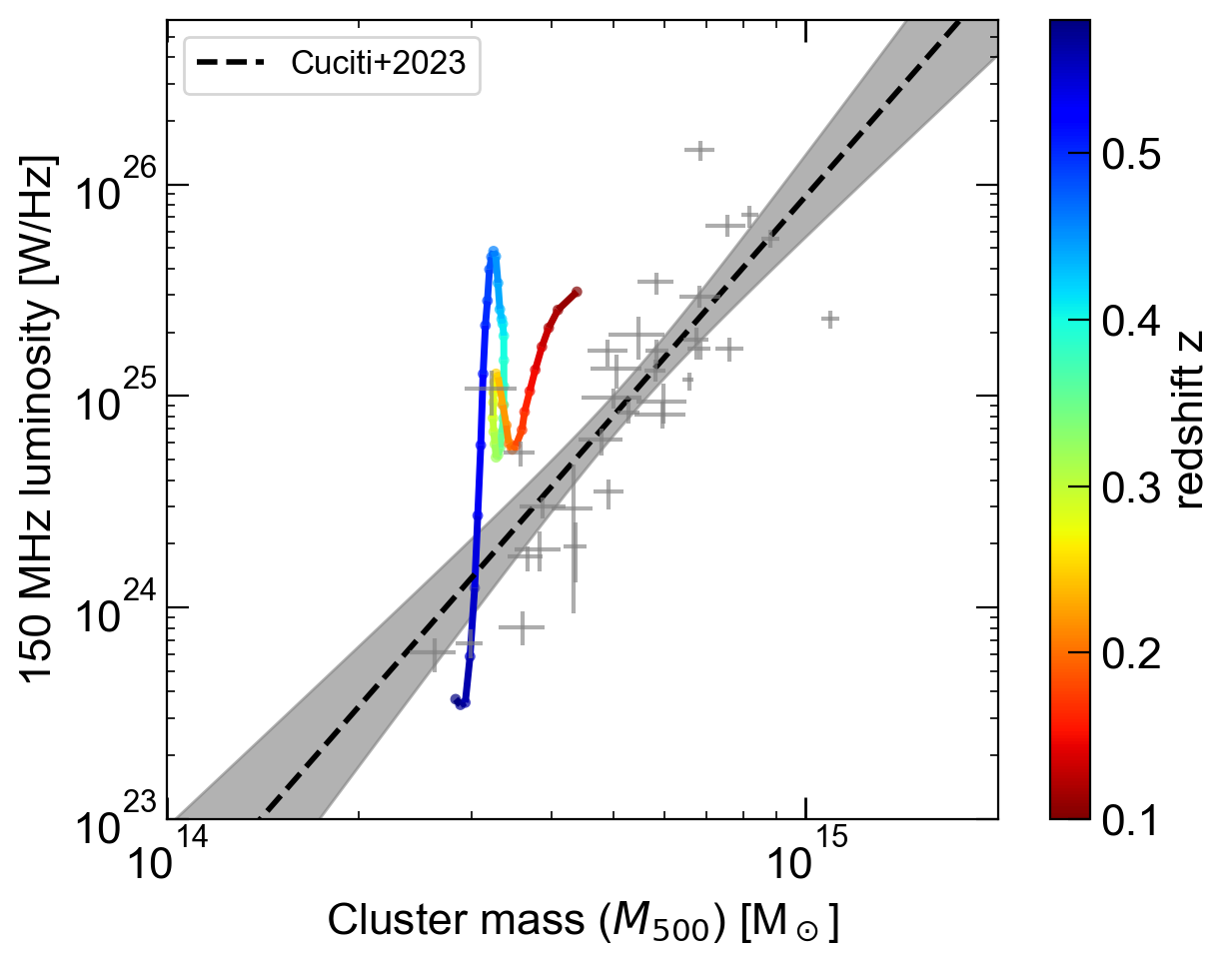}
    \caption{Evolution of the radio power at 150 MHz of the radio halo as a function of the halo mass $M_{500}$. The color scale indicates the redshift evolution in $0.6\leq z \leq0.1$. The data points and the correlation shown with the dashed line and the shaded region are adopted from \citet{Cuciti_2023}.}
    \label{fig:luminosity-mass}
\end{figure}

\subsection{Evolution of a radio halo \label{sec:halo}}

In this work, we show that the turbulent reacceleration model can reproduce several observational properties of the radio bridge using the parameters listed in Table~\ref{tab:params}.
As seen in Fig.~\ref{fig:bridge_radio_map_z}, our simulation produces not only the radio bridge but also halo-like emission around one of the clusters (on the left side of each panel).
Using the reacceleration model constrained by the observed properties of the radio bridge, we examine the synchrotron emission in the cluster region.
We note that the model adopted in this work is not intended to provide a fully self-consistent description of both the radio bridge and the radio halo.
In particular, the prescription for the CR injection (Sect.~\ref{sec:seed}) is simplified, and the contribution from secondary CR electrons is neglected.
With these caveats in mind, we briefly examine the evolution of the halo that forms in our simulation.
\par

We study the time evolution of the radio luminosity at 150 MHz by integrating the radio emissivity within a sphere of radius $r_{\rm halo} =R_{500}$, where $R_{500}$ is the radius enclosing a matter overdensity 500 times the critical mass density of the Universe.
The cluster center in three-dimensional space is defined as a local peak of the total matter (gas and dark matter) density, and $R_{500}$ is found by evaluating the overdensity as a function of radius.
In Fig.~\ref{fig:luminosity-mass}, we show the halo evolution in the correlation plot between the radio power $P_{150}$ and the cluster mass $M_{500}$, where $M_{500}$ is the total mass within the sphere of $r = R_{500}$.
The data points show the halo luminosities in the LOFAR survey \citep[][]{Cuciti_2023}.
Around $z \approx 0.5$, the cluster experiences a merger with another halo moving in $-X$ direction, and the mass increases from $M_{500} \approx 2.5 \times10^{14}$ M$_\odot$ at $z \approx 0.55$ to $M_{500} \approx 3.2 \times10^{14}$ M$_\odot$ at $z \approx 0.3$.
The radio power rapidly increases during the merger and reaches a peak of $P_{150}\approx10^{25.5}$ W Hz$^{-1}$ at $z\approx0.45$.
Although this peak value is almost 30 times larger than expected from the observed correlation, the power decays by a factor of $\sim 10$ within $\sim$800 Myr, from $z\approx0.45$ to $z\approx0.35$.
Between $z \approx 0.35$ and $z \approx 0.2$, both the radio power and the cluster mass remain nearly constant at $P_{150}\approx7\times10^{24}$ W Hz$^{-1}$ and $M_{500}\approx3\times10^{14}$ M$_\odot$.
\par

A similar evolutionary path of a simulated radio halo was reported by \citet{Donnert_2013}.
They studied the time evolution of radio luminosity in an idealized two-body merger system, considering turbulent reacceleration of CREs through transit-time-damping acceleration with fast modes \citep{BL07}.
Although a direct comparison between an idealized merger simulation and a cosmological simulation is not straightforward, we outline here the similarities and differences between our results and those of \citet{Donnert_2013}.

\citet{Donnert_2013} reported three phases of the evolution; infall phase, reacceleration phase, and decay phase.
In the first infall phase, they found the increase of radio luminosity by a factor of $\approx$30 within a few hundred Myr. In our simulation, the rapid evolution in $0.5\lesssim z\lesssim0.45$ corresponds to this phase. 
In the reacceleration phase, they found that the radio luminosity is sustained for several hundred Myr.
Similarly, we find that the halo maintains its luminosity in the range $0.4 \lesssim z \lesssim 0.2$, although the duration is longer by a factor of $\approx 2$ in our case.
We identify a small mass clump plunging into the cluster center around $z\approx0.3$, which might provide additional turbulent energy in the cluster.
Note that semi-analytic models for radio halo statistics assume a few Gyr for lifetimes of radio halos \citep[e.g.,][]{Cassano_Brunetti_2005,Nishiwaki25}.
Unlike \citet{Donnert_2013}, we do not see the gradual decaying phase of the radio halo. 
Instead, in our simulation the evolution after $z = 0.2$ is largely influenced by mass accretion from the inter-cluster filament and by shocks generated during the merger of the small clump with the filament (Sect.~\ref{sec:bridge}), which eventually leads to an increase in radio luminosity.
\par

The evolutionary track of radio halo in the luminosity--mass relation passes near several observed halos, particularly those at the high-luminosity end.
In addition, the trajectory aligns with the slope of the correlation during the evolution from $z\approx0.2$ to $z\approx0.05$.
The consistency between the observed relation and the reacceleration model adopted here should be investigated further through a systematic study of multiple halos formed in cosmological simulations.
\par

We note that the halo luminosity is expected to depend on the model parameters. 
We emphasize that the normalization factor $\phi$ has an uncertainty of approximately a factor of ten, due to the poorly constrained volume of the A399--A401 bridge and the choice of the simulated snapshot used for comparison (Sect.~\ref{sec:bridge_spect}). 
In addition, the halo power can vary by a factor of $\approx3$ depending on the adopted threshold for the CR injection, $\rho_{\rm thr}$, without inducing major changes in the luminosity and the spectrum of the bridge (see Sect.~\ref{sec:models}).

\par

\begin{table}[]
    \centering
    \caption{Additional models for the parameter study in Sect.~\ref{sec:models}.}
    \begin{tabular}{ccccc}
      \hline
      \hline
    model & $\eta_B$ & $\psi$ & $\rho_{\rm thr}$ & $\phi$  \\
    \hline
    fiducial & 0.05 & 0.5  & $10^{-28}$ g cm$^{-3}$ & $6\times10^{-8}$ \\
     1 & 0.05 & 0.5  & $10^{-27}$ g cm$^{-3}$ & $1.5\times10^{-6}$ \\
     2 & 0.05 & 0.5  & $10^{-29}$ g cm$^{-3}$ & $1.5\times10^{-8}$ \\
     3 & 0.02 & 0.5  & $10^{-28}$ g cm$^{-3}$ & $2.2\times10^{-8}$\\
     4 & 0.08 & 0.5  & $10^{-28}$ g cm$^{-3}$ & $9.0\times10^{-8}$ \\
     5 & 0.05 & 0.4  & $10^{-28}$ g cm$^{-3}$ & $5.0\times10^{-9}$\\
     6 & 0.05 & 0.6  & $10^{-28}$ g cm$^{-3}$ & $7.3\times10^{-7}$\\
     \hline
    \end{tabular}
    \tablefoot{We fix $L = 150$ kpc and $z_i = 1$ in all of the models.}
    \label{tab:models}
\end{table}

\begin{figure}
    \centering
    \includegraphics[width=\linewidth]{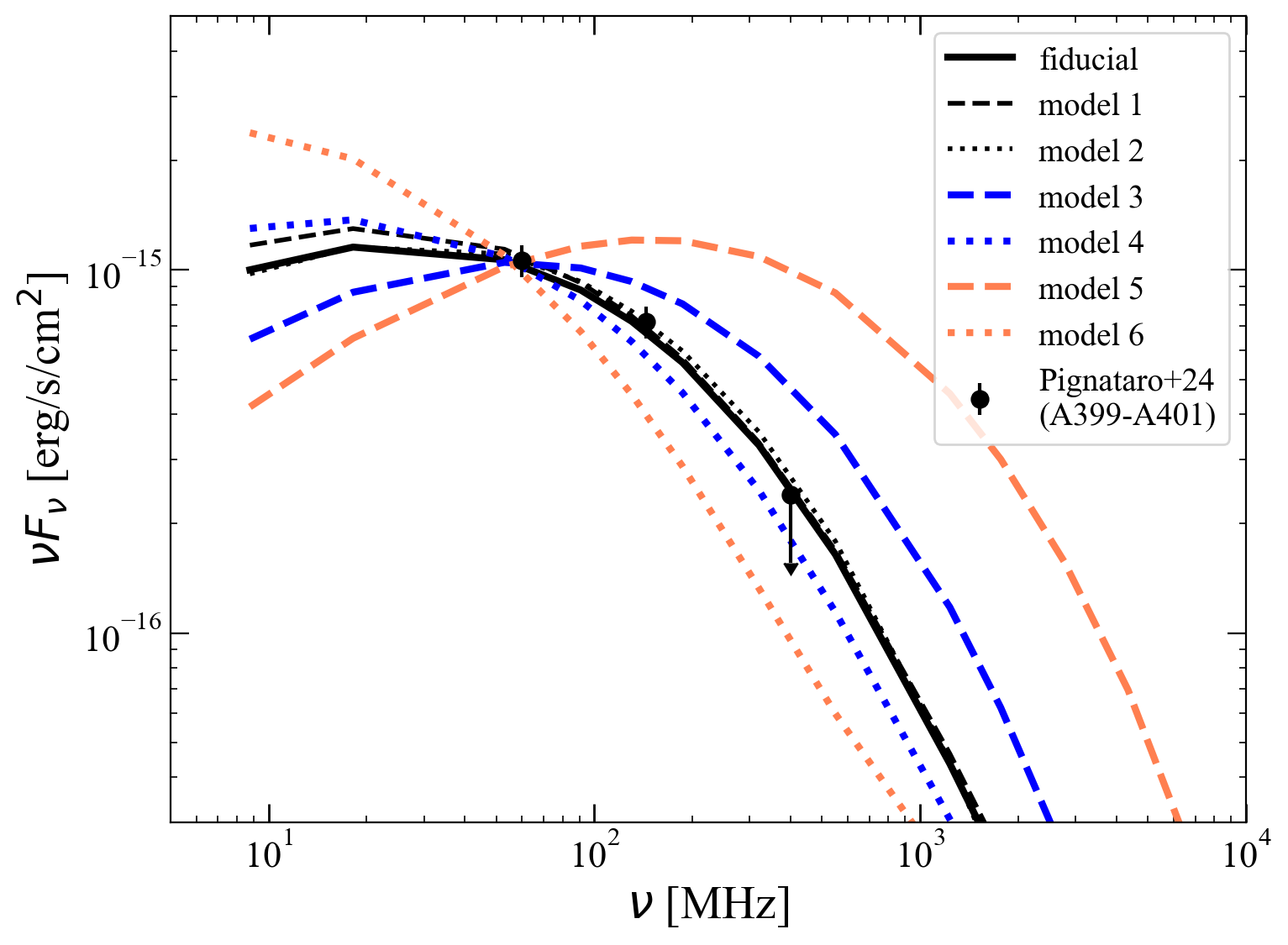}
    \caption{Spectra of the radio bridge for various models. The solid black line shows the fiducial model presented in Sect.~\ref{sec:results}. Models 1 and 2 yield spectra that are similar to the fiducial one, shown by the dashed and dotted black lines, respectively. The blue curves correspond to $\eta_B = 0.02$ (model 3; dashed) and $\eta_B = 0.08$ (model 4; dotted), while the orange curves show the cases with $\psi = 0.4$ (model 5; dashed) and $\psi = 0.6$ (model 6; dotted). For comparison, all spectra are normalized to match the 60 MHz radio power of A399--A401. The model parameters are summarized in Table~\ref{tab:models}.}
    \label{fig:spect_models}
\end{figure}

\subsection{Parameter study \label{sec:models}}

In this section, we explore how the simulated radio emission is affected by variations in the model parameters.
Our model behaves similarly to that of \citet{Beduzzi_2024}, as both adopt the same reacceleration mechanism (Eq.~(\ref{eq:Dpp})) and share several parameters.
As discussed in \citet{Beduzzi_2024}, the radio emissivity scales linearly  with the normalization parameter $\phi$, and the redshift of the initial seeding, $z_i$, has only a minor effect unless it is set close to the observation redshift ($z=0.1$).
Here, we focus on three parameters: $\rho_{\rm thr}$, $\eta_B$, and $\psi$.
The threshold density for the initial CR seeding, $\rho_{\rm thr}$, is introduced in this work.
The parameters $\eta_B$ and $\psi$ have significant impacts on the magnetic field and the reacceleration efficiency.
We construct six models, summarized in Table~\ref{tab:models}; in each case, only one parameter is varied with respect to the fiducial model (Table~\ref{tab:params}).
The normalization parameter $\phi$ is determined for each model by fitting the 60 MHz flux of the A399--A401 radio bridge.
Models 1 and 2 explore larger and smaller values of $\rho_{\rm thr}$.
Models 3 and 4 probe the dependence on the dynamo efficiency, $\eta_B$ (Sect.~\ref{sec:dynamo}).
Models 5 and 6 vary the parameter $\psi$, which represents the normalized mean free path of CREs and strongly influences the efficiency of turbulent reacceleration (Sect.~\ref{sec:reacc}).
\par

The parameters $\eta_B$ and $\psi$ affect the evolution of the CR spectrum along tracer trajectories and therefore require the FP simulation to be rerun for models 3, 4, 5, and 6.
To reduce the computational cost, we select $N = 5\times10^5$ tracers residing in the bridge region (white box in Fig.~\ref{fig:Enzo_map_z}) and compute the emission exclusively from these tracers.
Although this prevents us from studying changes in the full projected radio map, it still allows us to compute the integrated synchrotron spectrum in the bridge.
On the other hand, as explained in Sect.~\ref{sec:seed}, the effect of $\rho_{\rm thr}$ can be evaluated in post-processing, so we do not need to rerun the simulation for models 1 and 2.
\par

Fig.~\ref{fig:spect_models} shows the radio spectra for the models introduced above. 
The solid black line corresponds to the fiducial model presented in Sect.~\ref{sec:results}.
We find that $\rho_{\rm thr}$, tested in models 1 and 2, does not strongly affect the shape of the synchrotron spectrum (dashed and dotted black lines).
In model 1, the inferred value of $\phi$ is larger than in the fiducial case because a larger threshold density for CR seeding leads to fewer CREs in the bridge region.
We also find that the relative radio brightness of the halo and bridge components depends on $\rho_{\rm thr}$.
As shown in Fig.~\ref{fig:bridge_profiles}, the halo emission is brighter in model 1 ($\rho_{\rm thr} = 10^{-27}$ g cm$^{-3}$) than the fiducial model ($\rho_{\rm thr} = 10^{-28}$ g cm$^{-3}$) by a factor of $\approx$4.
However, we caution that the brightness profile evolves with time and also depends the other parameters, such as $z_{\rm i}$, $\eta_B$, and $\psi$.
The connection between the halo and the bridge should be further investigated in future work with an improved CR injection model.
\par

The results for models 3 and 4 are shown with dashed and dotted blue lines in Fig.~\ref{fig:spect_models}, respectively. 
In both models 3 and 4, the typical magnetic field is much smaller than $B_{\rm cmb}(z) \approx 3.2(1+z)^4$ $\mu$G\footnote{The mean magnetic field in the bridge region becomes $\langle B\rangle\approx 0.28$ $\mu$G for $\eta_B = 0.02$ and $\langle B\rangle\approx 0.57$ $\mu$G for $\eta_B = 0.08$.}, so the radiative cooling of CREs is dominated by inverse-Compton losses.

In this regime, the radiative cooling of CRE at a given momentum $p$ is almost independent on the magnetic field value, because the IC losses dominate the synchrotron losses.
The magnetic field affects the reacceleration efficiency and the synchrotron emissivity.
Since $D_{pp} \propto \eta_B^{-1/2}$ (Sect.~\ref{sec:reacc}), the reacceleration becomes more efficient and produces a flatter CRE spectrum.
On the other hand, for a fixed CRE spectrum, a smaller magnetic field leads to a dimmer and steeper synchrotron spectrum.
In our model, however, the former effect dominates over the latter, and therefore a smaller (larger) $\eta_B$ leads to a flatter (steeper) synchrotron spectrum, as seen in Fig.~\ref{fig:spect_models}.
This trend has been noted also in \citet{Nishiwaki24}.

\par

More quantitatively, the above trend can be understood by comparing the timescales of reacceleration and radiative cooling.
Considering the synchrotron emission at a given observing frequency, $\nu_{\rm o}$, the cooling time of CREs, whose characteristic synchrotron frequency matches $\nu_{\rm o}$, scales with $t_{\rm cool}(\nu_{\rm o}) \propto \eta_B^{1/4}\nu_{\rm o}^{-1/2}$, as shown in Eq.~(6) of \citet{Nishiwaki24}.
The ratio of the reacceleration time and the cooling time scales as $t_{\rm acc}/t_{\rm cool}(\nu_{\rm o})\propto\eta_B^{1/4}\nu_{\rm o}^{1/2}$. 
For a fixed $\nu_{\rm o}$, this ratio becomes smaller for smaller $\eta_B$, leading to a flatter synchrotron spectrum.
In other words, the cut-off frequency of the synchrotron spectrum, which reflects the balance between $t_{\rm acc}$ and $t_{\rm cool}$, shifts to higher frequencies for smaller $\eta_B$.

\par

In summary, our reacceleration model can generate a wide range of spectral shapes within physically reasonable parameter values.
Future high-frequency observations will be essential to constrain the magnetic field strength and reacceleration efficiency in the bridge region.

\subsection{Limitations and caveats \label{sec:limitation}}

We aim to investigate the origin of radio bridges in pre-merger systems with a projected extension of a few Mpc, such as A399--A401 or A1758.
Using the simulation setup of \citet{Govoni_2019}, we assume that the 3D separation of clusters is about $\approx3-4$ Mpc at $z\approx0.1$. 
The resulting emission depends on the 3D separation and the evolutionary stage of the simulation, as discussed in Sects.~\ref{sec:bridge} and \ref{sec:bridge_spect}.
However, we note that the 3D geometry of the observed systems can differ from the simulated one due to the uncertainty in the extension along the LOS. 
For example, some authors suggested that the merger axis of the A399--A401 system may have a large inclination with respect to the plane of the sky \citep[e.g.,][]{Hincks_2022}.
\par

As discussed in Sect.~\ref{sec:bridge}, the turbulence in the radio bridge formed at $z\approx 0.1$ is clearly affected by the collision of the mass clump that occurred at $z\approx0.2$.
\citet{BV20} also noted that radio bridges could be powered by substructures orbiting the region connecting the two clusters.
The radio bridge might not have been generated without that event, although this cannot be demonstrated in the present work. A systematic study using a large box of a cosmological simulation would be valuable to investigate the occurrence of radio bridges before the core passage of cluster pairs.

Concerning the initial seeding of CREs, we assume a highly simplified scenario in which CREs are injected only once at a single epoch of the simulation, $z_{\rm i}$. 
We further assume that the number of CREs is directly proportional to the local thermal electron number. 
Recent work by \citet{Vazza_2025} investigated the CR seeding in the cosmic web and reported that CR spatial distribution roughly traces that of thermal gas, supporting this scaling assumption.
Although we find this prescription results in the generation of a radio bridge and has the capability to generate a radio halo simultaneously, there are some caveats; the results depends on the injection threshold, $\rho_{\rm thr}$ (Sect.~\ref{sec:models}), the scaling of Eq.~(\ref{eq:phi}) needs to be tested considering the history of AGN and galactic feedback in the ICM.
In reality, CRs can be supplied multiple times by AGN jets during the evolution of GCs. 
\par

In this exploratory work, we focus on the reacceleration of CREs in the filament region between a pair of merging clusters. 
Although a number of shocks are generated by the infall of mass clumps in the filament, the effect of shock (re)acceleration is not considered in this work. 
\citet{Govoni_2019} discussed the shock acceleration scenario as a possible mechanism for the radio bridge and concluded that the reacceleration at weak shocks can explain the observed morphology under peculiar conditions (see Sect.~\ref{sec:intro}).
However, the efficiency of CR acceleration at such weak shocks is still under debate and may depend on the Alfv\'enic Mach number and the obliquity of the magnetic field \citep[e.g.,][]{Kang_2007,Bykov_2019,Ryu_2019,Gupta_2025}.
Cosmological MHD simulations with spectrally-resolved CR modeling provide an opportunity to investigate the phenomenological constraints on the efficiency of shock acceleration in the cluster environment \citep[e.g.,][]{Boess_2023,Vazza_2023,Boss_2025,Ivleva_2026}.
Our tracer method onboard Enzo can be extended to include the on-the-fly shock finder to explore the CR acceleration at shocks.

\par

Hadronic interactions of CR protons can provide relativistic electrons in the filaments \citep[e.g.,][]{Brown_2017}. 
Unlike CREs, which are subject to severe energy losses, CR protons can survive for a cosmological time in the inter-galactic medium and continuously produce secondary particles. 
Within the framework of the turbulent reacceleration model, the spectrum of secondary electrons can be non-trivial, since the reacceleration can also act on CRPs \citep[e.g.,][]{Nishiwaki25}. 
Fig.~\ref{fig:bridge_spectrum} shows the similarity of the radio spectrum obtained in this work and that of \citet{BV20}.
Considering that \citet{BV20} is based on a MHD snapshot and ignores the long-term evolution of the system, this similarity possibly stems from the short timescales of CRE cooling and reacceleration compared with the dynamical time of the merger system. 
Our tracer-particle method is suitable for studying the long-term evolution of the CRP spectrum. 
In future work, we will investigate the evolution of CR protons and secondary electrons in the cosmic filaments.
\par

Finally, we note that the initialization of magnetic fields in our simulation implicitly assumes a primordial generation mechanism for extragalactic magnetic fields, in which a volume-filling magnetic field seed ($10^{-10} \rm G$ comoving) is already present at the initial redshift of the simulation ($z=30$). 
While the origin of cosmic seed magnetic fields remains debated \citep[e.g.,][and references therein]{sub16,vach21,va21magcow}, recent statistical modeling of extragalactic Faraday Rotation and synchrotron radio emission from filaments in the cosmic web appear to prefer primordial large-scale magnetic fields with $\sim 0.1-0.4 \rm nG$ \citep[e.g.,][and references therein]{2025Univ...11..164C}, i.e., in line with our assumption.

The local turbulent amplitude recorded by the tracer particles was used to estimate the dynamo-amplified magnetic field and the reacceleration efficiency.
We produced simulated radio intensity maps by integrating the synchrotron emissivity calculated for each tracer along the LOS.
Our main findings can be summarized as follows:

\section{Conclusions \label{sec:conclusion}}
An increasing number of observations report diffuse radio emission in cosmic filaments connecting merging clusters or groups.
The mechanism responsible for accelerating CR particles in such low-density environments remains elusive.
In this work, we extend previous studies using advanced numerical simulations to investigate the turbulent reacceleration of CREs in the inter-cluster region and the formation of a radio bridge.
Using the AMR code Enzo, we conduct a cosmological MHD simulation of a system similar to A399--A401, which is known to host a 3 Mpc-long radio bridge.
We upgrade the tracer method onboard Enzo and generate $N = 1.5\times10^7$ particles to sample the baryonic mass of $M_{\rm gas} = 3\times10^{14}$ M$_\odot$.
We assume a population of CREs associated with each tracer particle and follow their spectral evolution with our parallel FP solver.

Based on the turbulent reacceleration model of \citet{BV20}, our methodology enables us to follow the evolution of CREs over cosmological time. This allows a more detailed characterization of the bridge emission and enables a direct and quantitative comparison with current observations. In particular, we present (i) spatially resolved radio maps of the bridge (Fig.~\ref{fig:bridge_radio_map_z}), (ii) the temporal evolution of the merging system and the emergence of the radio bridge (Fig.~\ref{fig:bridge_evolution}), (iii) spatially resolved spectral index maps (Fig.~\ref{fig:spix}), (iv) detailed spatial profiles across the bridge (Fig.~\ref{fig:bridge_profiles}), and (v) a quantitative radio–X-ray point-to-point correlation analysis (Fig.~\ref{fig:point_point}).
Our main findings in this work can be summarized as follows:
\par

\begin{enumerate}
    \item Our run-time tracer method implemented in Enzo provides a more accurate reproduction of the matter distribution in the original MHD simulation than the post-processing methods used in previous works (Appendix~\ref{app:trc_map}). 
    \item The simulation generates diffuse emission between merging clusters that resembles radio bridges (Fig.~\ref{fig:bridge_radio_map_z}).
    \item The collision of a small mass clump at $z \approx 0.15$ induces significant turbulence in the inter-cluster region, which appears to trigger the formation of the radio bridge. This turbulence is further amplified during the subsequent compression of the filament due to the approaching motion of the two clusters (Fig.~\ref{fig:bridge_evolution}).
    \item The gas density in the bridge region is about 1/10 of that in the cluster region, whereas the turbulent power density differs only by a factor of $\approx 2$. The reacceleration is efficiently working in the bridge region due to the high turbulent Mach number ($M_{\rm s}\approx0.3-0.4$).
    \item The properties of the simulated bridge emission are consistent with observations. In particular, we show that the observed radio spectrum of the bridge in A399--A401 (Fig.~\ref{fig:bridge_spectrum}) and the radio--X and radio--SZ correlations (Fig.~\ref{fig:point_point}) can be reproduced with our reference model parameters ($\eta_B = 0.05$ and $\psi = 0.5$). Depending on the model parameters and evolutionary stage, the simulated emission shows a variety of steep spectra (Fig.~\ref{fig:spect_models}).
    \item This study explores whether the turbulent reacceleration model can serve as a viable formation mechanism for radio bridges. We emphasize several simplifications in our model, including the treatment of CR injection, the neglect of secondary electrons, and the use of magnetic field estimated in post-processing.
   
\end{enumerate}

An increasing number of studies have combined high-resolution MHD simulations with spectrally resolved CR modeling to investigate diffuse emission in galaxy clusters and beyond \citep[e.g.,][]{Pinzke_2017,Vazza_2023,Boess_2023,Beduzzi_2024,Dominguez-Fernandez_2026}.
In this context, we have improved an existing tracer-based method applicable to Enzo simulations and investigated CR reacceleration in radio bridges.

Future studies will enable us to investigate more detailed physical processes related to diffuse synchrotron emission in the cosmic web, including comparisons among different reacceleration models and a self-consistent modeling of both radio halos and radio bridges.

\begin{acknowledgements}
We thank the anonymous referee for the careful review of the manuscript and for the constructive comments and suggestions, which helped improve the quality and clarity of this paper.
This work has received funding from the European High Performance Computing Joint Undertaking (JU) under grant agreement No 101093441, supported by contributions from multiple member states. The authors would like to thank all project partners and collaborators for their contributions and ongoing commitment to the success of the SPACE project.
K.N. and G.B. acknowledge financial support from INAF grant "Theory and simulations of nonthermal phenomena in galaxy clusters and beyond."
F.V. acknowledges funding under
    the European Union’s  Horizon Europe program through the ERC Synergy Grant COSMOMAG (Project Id. 101224803). 
We acknowledge INAF for awarding this project access to the LEONARDO supercomputer, owned by the EuroHPC Joint Undertaking, hosted by CINECA (Italy) and the LEONARDO consortium.
The computations were conducted using the resources allocated to the projects designated "INA24C5B12" and "INA24C6B13," with K.N. serving as the Principal Investigator.  F.V. acknowledges the usage of online storage tools kindly provided by the INAF Astronomical Archive (IA2) initiative (http://www.ia2.inaf.it). 
Numerical computations were in part carried out on PC cluster at the Center for Computational Astrophysics, National Astronomical Observatory of Japan.
\end{acknowledgements}

\bibliographystyle{aa}
\bibliography{ref}

\begin{appendix} 
\section{Accuracy test of Fokker-Planck solver}\label{app:FP}
\begin{figure*}
    \centering
    \begin{minipage}{0.48\linewidth}
        \includegraphics[width=\linewidth]{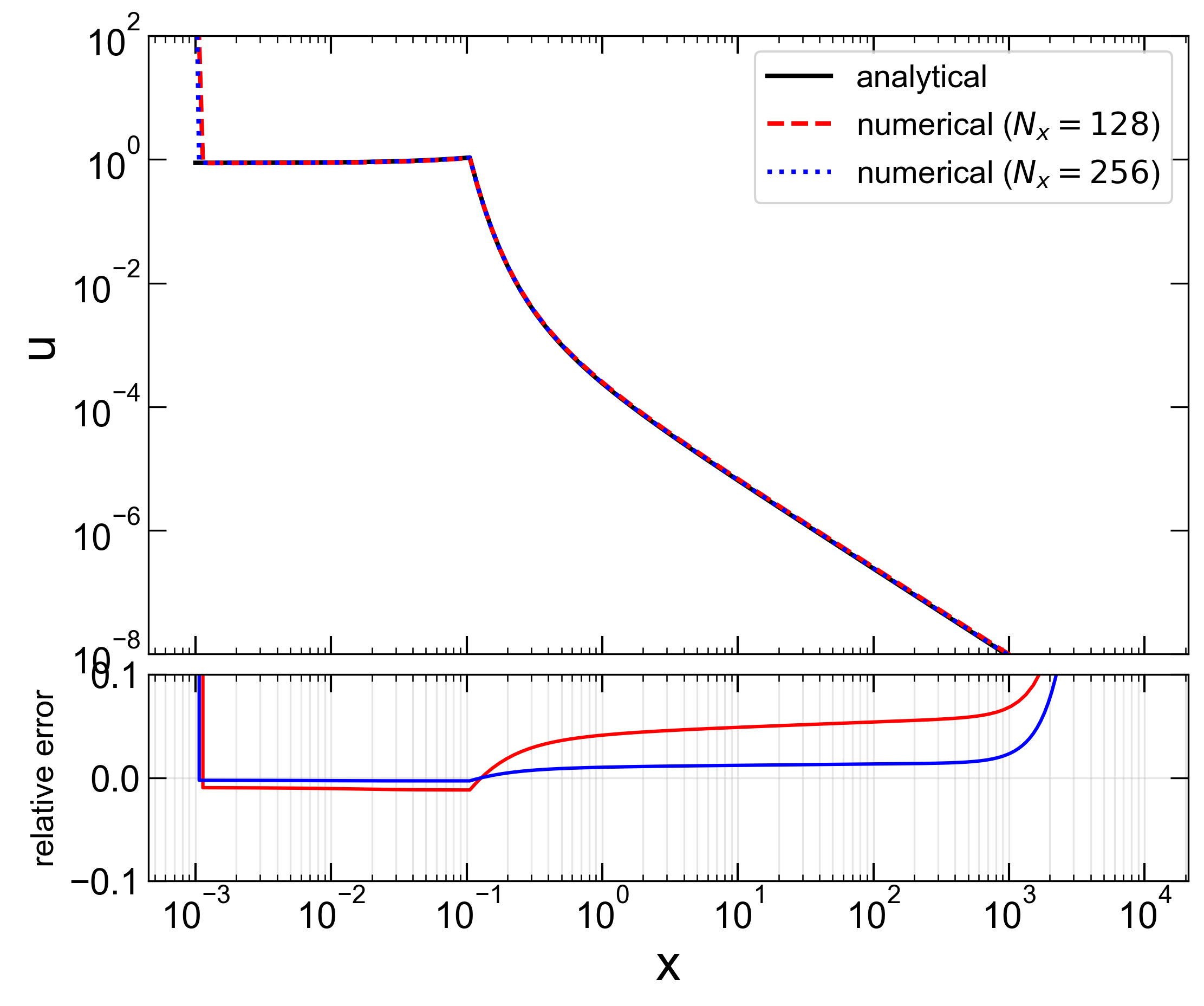}
    \end{minipage}
    \begin{minipage}{0.48\linewidth}
        \includegraphics[width=\linewidth]{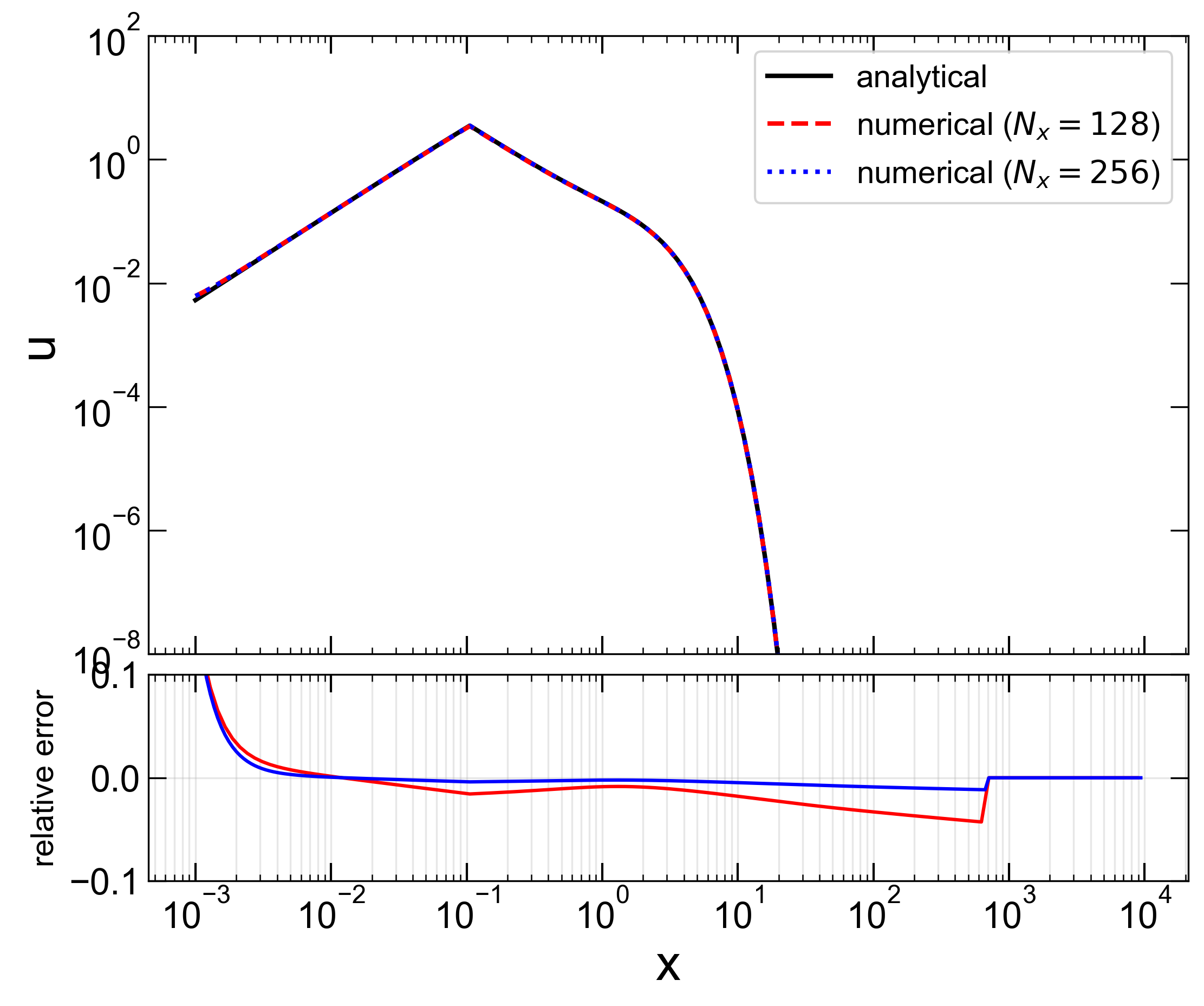}
    \end{minipage}
    \caption{
    Comparison to the analytical solution. Left panel: case for $B(x) = -x+1$. Right panel: case for $B(x) = -x+x^2$. The black solid lines show the analytical solutions. The red and blue lines show numerical solutions for different resolutions for $x$ bin: $N_x = 128$ and 256, respectively. In the lower panels, we show the relative error of the numerical solution to the analytical one.
    }
    \label{fig:FP_test}
\end{figure*}

In the FP solver we developed in \citet{Nishiwaki25}, we apply the finite-difference scheme developed by \citet{Chang_Cooper_70}.
The accuracy of this method was examined by \citet{Park_Petrosian_96} (hereafter PP96).
Following \citet{Donnert_Brunetti_2014}, we test our FP solver by repeating the benchmark test presented in PP96 and discuss the accuracy of the method.
\par

Our FP solver operates by specifying the coefficients in the following equation (Eq.~(1) of PP96),
\begin{equation}
    \frac{\partial u}{\partial t} = \frac{1}{A(x)}\frac{\partial}{\partial x}\left[C(x)\frac{\partial u}{\partial x} + B(x)u\right] - \frac{u}{T(x)} + Q(x),
    \label{eq:FP_test}
\end{equation}

where $A(x) \equiv 1$ is the phase factor, $B(x)$, $C(x)$, $T(x)$, and $Q(x)$ are the advective, diffusive, sink, and source terms, respectively.
In these test calculations, we consider a hard-sphere-type diffusion, $C(x) = x^2$, a uniform sink $T(x) \equiv 1$, and a delta-function source $Q(x) = \delta(x-x_{\rm inj})$.
The reacceleration mechanism considered in this work, as many of second-order Fermi acceleration models, has a hard-sphere form ($D_{pp} \propto p^2$, Eq.~(\ref{eq:Dpp})).
We test two cases for $B(x)$; $B(x) = -x + 1$ and $B(x) = -x + x^2$. The former corresponds to the case with a constant shift in $x$, similar to the adiabatic expansion, while the latter has a cooling proportional to $x^2$, similar to radiative cooling.
For these cases, we can directly compare the numerical solutions to the analytical solutions given by \citet{Park_Petrosian_95}.

\par

We set $N_x=128$ discretized bins for $x$ in the range of $x_{\rm m}\leq x \leq x_{\rm M}$, where $x_{\rm m} = 10^{-3}$ and $x_{\rm M} = 10^{4}$. 
The bins are equally spaced in the logarithmic scale. 
We note that the adopted resolution (128 bins over seven dex) is the same as that used for the CRE momentum bin in Sect.~\ref{sec:FP}.
The fully implicit discretization of Eq.~(\ref{eq:FP_test})
leads to a tridiagonal system of linear
equations:
\begin{equation}
    -a_{m}u_{m-1}^{n+1} + b_{m}u_{m}^{n+1} -c_{m}u_{m+1}^{n+1} = r_m,
\end{equation}
where the subscript $m$ and superscript $n$ denote the $x$ bin and the time step, respectively.
We impose a no-flux boundary condition, $a_0 = 0$ and $c_M = 0$.
The coefficients $a_{m}$, $b_{m}$, $c_{m}$, and $r_{m}$ for the Chang-Cooper method are presented in PP96.
We solve the tridiagonal system using a direct Gaussian elimination scheme.
The initial condition is set to $u(x)=0$ at $t= 0$. We adopt $\Delta t = 10^{-3}$ for the time step and follow the solution up to $t = 10$.

\par

Figure~\ref{fig:FP_test} compares the numerical solutions at $t = 10$ with the exact solution by \citet{Park_Petrosian_95} (their Eqs.~(71) and (72)) for $B(x) = -x + 1$ (left panel) and $B(x) = -x + x^2$ (right panel).
The red and blue lines show the results for $N_x = 128$ and 256, respectively.
The lower panels show the relative errors with respect to the analytical solution, $(u_{\rm num} - u_{\rm ana})/u_{\rm ana}$.
We recover the results of PP96: the numerical solution matches the analytical one well in wide range of $x$.
For $B(x) = -x + x^2$, the typical relative error around $x\sim1$, where acceleration balances with cooling, is a few percent. 
As already pointed out by PP96, the error increases toward the boundary, and there is a jump in a single bin $x=x_{\rm m}$ in the case of $B(x) = -x + 1$ (left panel).
These errors neither grow nor propagate with time after the steady-state solution has been achieved.
The solutions at $t=100$ do not show any noticeable differences compared to those at $t=10$.
The error involved in the FP solver is negligible considering the variation of the results due to the uncertainties in the model parameters (Sect.~\ref{sec:method}).
It is also evident that the accuracy of the solver improves with increasing momentum resolution, albeit at the expense of increased computational cost.
These tests ensure that the Chang-Cooper scheme of PP96 is correctly implemented in our FP solver.

We note that \citet{Donnert_Brunetti_2014} adopted a different boundary condition, whereas we apply the no-flux boundary condition following PP96. 
In their approach, the spectrum is truncated near the boundaries and the numerical solution is extrapolated beyond the computational domain. 
While this method can suppress the unphysical pile-up of particles at the boundary (see Fig.~\ref{fig:FP_test}, left panel), it may introduce systematic uncertainties in the boundary flux due to errors in the extrapolation.
In practice, for the case of CREs in galaxy clusters and filaments, the accumulation of particles at the minimum momentum $p_{\rm min}$ may lead to an unphysically large flux when acceleration dominates over Coulomb cooling at $p_{\rm min}$. 
We circumvent this issue by setting $p_{\rm min}/m_ec = 0.1$, at which the Coulomb cooling timescale becomes shorter than $10\,{\rm Myr}$ even in extremely low-density regions with $n_{\rm gas} \sim 10^{-5}\,{\rm cm^{-3}}$, and remains shorter than the timescales of adiabatic compression and turbulent reacceleration.

\section{Tracer generation method}\label{app:trc_generation}
We study the trajectory of the baryonic fluid using Lagrangian tracer particles.
The tracer particles used in this work are generated and propagated during the runtime of the Enzo simulation, in contrast to previous studies that employed post-processing tracers.
We modified the preliminary tracer implementation in the public version of Enzo 2.6 so that one can specify the spatial region and the epoch of tracer generation, in addition to the tracer mass resolution.
We introduce new parameters to control the redshift of tracer generation, the left and right boundaries of the sampling region, and the minimum and maximum tracer masses ($M_{\rm trc}^{\rm min}$ and $M_{\rm trc}^{\rm max}$).

\par

In our setup, tracers are generated at $z = 1$ in the $(20\,{\rm Mpc})^3$ box encompassing the merger system composed of two massive clusters.
This system has a total (baryon + dark matter) mass of $\sim10^{15}$ M$_\odot$.
Assuming a baryon fraction of $f_{\rm bar}\sim 0.1$ \citep[e.g.,][]{Allen_2008,Gonzalez_2013}, the total baryonic mass sampled by the tracers is $M_{\rm trc}^{\rm tot}\sim10^{14}$ M$_\odot$.
The tracer mass resolution is controlled by specifying the allowed mass range, defined by $M_{\rm trc}^{\rm min}$ and $M_{\rm trc}^{\rm max}$.
In this work, we adopt $M_{\rm trc}^{\rm min} = 1.5\times10^{7}$ M$_\odot$ and $M_{\rm trc}^{\rm max} = 1.5\times10^{8}$ M$_\odot$.
\par

In the original Enzo 2.6 implementation, tracers are generated only on the top-level grid with the coarsest resolution.
We modified this implementation to allow particle generation on the finest AMR level available at each spatial position to follow the highest-resolution representation of the gas field.
Specifically, the code searches through all grids and cells within the injection region specified by the input parameters and identifies cells that are not covered by any finer-level grid. 
The tracer particles are generated only in these flagged cells, i.e., cells that are not further refined by any child grid.
This approach avoids duplicating particles in overlapping AMR regions.
\par

When tracer particles are generated, the code measures the baryon mass of each cell as $M_{\rm cell} = \rho_{\rm gas}(\Delta x/(1+z))^3$, where $\Delta x$ is the comoving cell width. 
The computed cell mass is then compared with $M_{\rm trc}^{\rm min}$ and $M_{\rm trc}^{\rm max}$.
In low-resolution cells, the cell mass can exceed $M_{\rm trc}^{\rm max}$. 
In this case, we generate $\left\lfloor \frac{M_{\rm cell}}{M_{\rm trc}^{\rm max}} \right\rfloor$ tracers with mass $M_{\rm trc} = M_{\rm trc}^{\rm max}$, and, if the residual mass exceeds $M_{\rm trc}^{\rm min}$, one additional tracer with the residual mass $M_{\rm trc} = M_{\rm cell} - M_{\rm trc}^{\rm max}*\left\lfloor \frac{M_{\rm cell}}{M_{\rm trc}^{\rm max}} \right\rfloor$.
When the cell mass falls between $M_{\rm trc}^{\rm min}$ and $ M_{\rm trc}^{\rm max}$, a single tracer with $M_{\rm trc} = M_{\rm cell}$ is generated.
When $M_{\rm cell} < M_{\rm trc}^{\rm min}$, the tracer mass is too large to represent the cell. 
In this case, we adopt a Monte-Carlo sampling approach to include the contribution of such low-mass cells in a statistical sense: a tracer particle with $M_{\rm trc} = M_{\rm trc}^{\rm min}$ is generated with probability of $P = M_{\rm cell}/M_{\rm trc}^{\rm min}<1$. 
Although this method smooths the small-scale mass distribution below the mass resolution $M_{\rm trc}^{\rm min}$, it conserves the baryonic mass statistically.
For each tracer, three random numbers in the range $[0,1)$ are generated to assign its initial $(x,y,z)$ position within the host cell.
Tracer particles are treated as passive (virtual) particles and do not contribute to the calculation of the gravitational potential or the MHD evolution.
\par

\begin{figure}
    \centering
    \includegraphics[width=\linewidth]{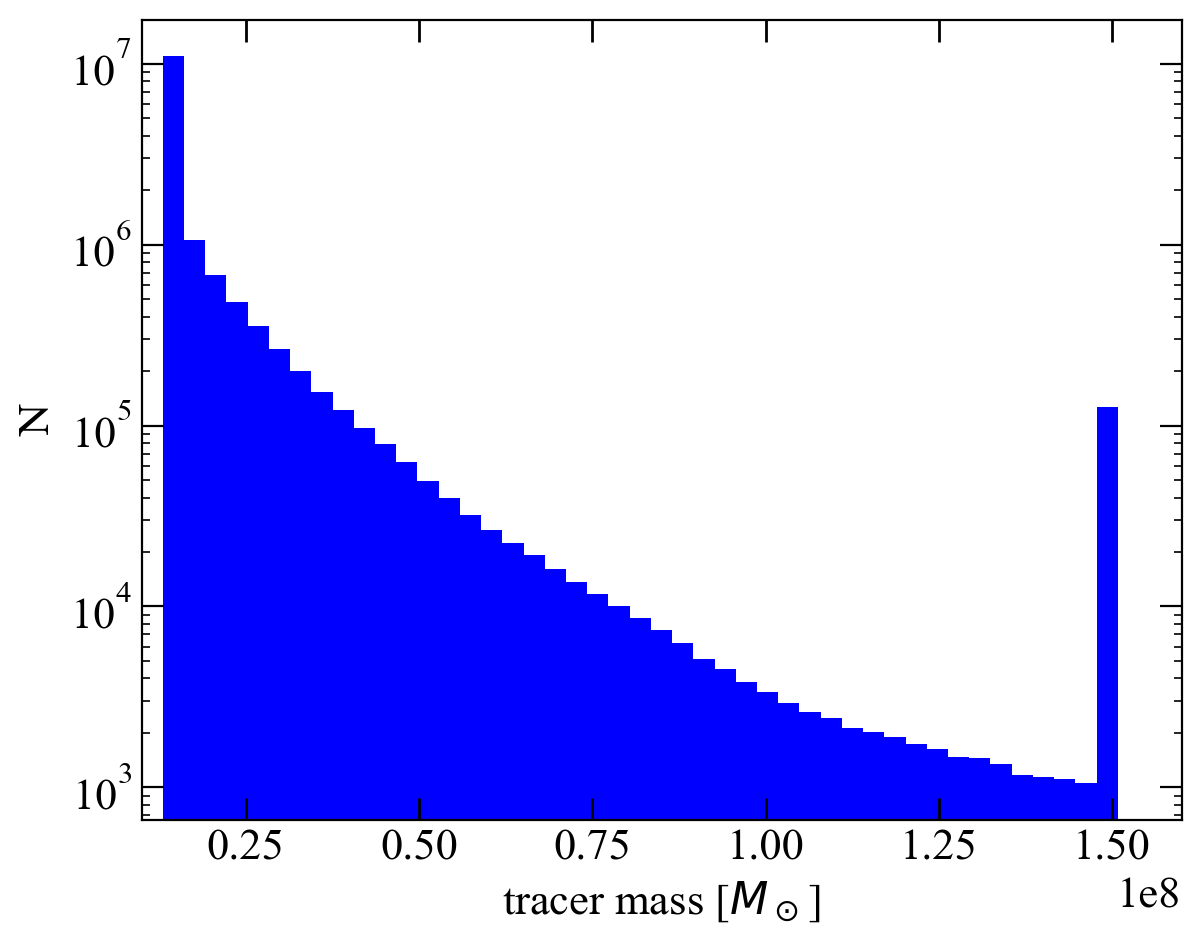}
    \caption{
    Mass distribution of tracers used in this work. The total number of tracers is $N_{\rm tot} = 1.5\times 10^{7}$. The baryonic mass sampled by the tracers is $M^{\rm tot}_{\rm trc} = 3.0\times10^{14}$ M$_\odot$. 
    }
    \label{fig:tracer_mass}
\end{figure}

In Fig.~\ref{fig:tracer_mass}, we show the distribution of tracer masses used in this work. 
Our method generates $N_{\rm tot} = 1.5\times 10^{7}$ tracers, sampling a baryonic mass of $M^{\rm tot}_{\rm trc} = 3.0\times10^{14}$ M$_\odot$. Approximately 70\% of the tracers have the masses equal to $M_{\rm trc}^{\rm min}$, while 0.8\% have $M_{\rm trc} = M_{\rm trc}^{\rm max}$. 
The effective mass resolution of the tracers can be estimated as $M_{\rm trc}^{\rm eff}\equiv \sum_i^{N_{\rm trc}}(M_{\rm trc,i})^2/\sum_i^{N_{\rm trc}}M_{\rm trc,i}\approx3\times10^{7}$ M$_\odot$.
This value is not uniform throughout the simulation volume and tend to be smaller in higher density regions.

\section{Tracer deposition method \label{app:trc_deposit}}
For a visualization purpose, we deposit the tracer quantities to a three-dimensional mesh with a fixed resolution.
The most straightforward way would be to assign each tracer to its nearest grid cell.
However, due to the limited mass resolution of the tracers, this method is severely affected by shot noise when the mesh size is comparable to (or less than) the virtual size of the tracer particle.
To minimize the error in the reproduced mesh, we take into account the overlap between the tracer particle and the surrounding cells.
\par

For simplicity, we assume that each tracer has a spherical shape.
The effective radius of the tracer is therefore given by
\begin{equation}
    R_{\rm trc} = \left( \frac{3 V_{\rm trc}}{4\pi} \right)^{1/3}.
\end{equation}

Using the tracer mass and the local gas density, we assign each particle a finite volume $V_{\rm trc}$ calculated as 
\begin{equation}
    V_{\rm trc} = \frac{M_{\rm trc}}{\rho_{\rm gas}},
\end{equation}
where $M_{\rm trc}$ is the tracer mass and $\rho_{\rm gas}$ is the gas density.
For example, a tracer particle representing a gas mass of $M_{\rm trc} \approx 2 \times10^7$ M$_\odot$ in the bridge region with $\rho_{\rm gas} \approx 2\times10^{-28}$ g cm$^{-3}$ occupies the volume of $\approx$(19 kpc)$^3$.
For simplicity, we assume that each particle has a spherical shape.
The effective radius is given by $R_{\rm trc} = \left(\frac{3}{4\pi}V_{\rm trc}\right)^{1/3}$.
We compute the geometric overlap between the sphere of the particle and the surrounding mesh cells based on the central position of the particle.
Cells that intersect the sphere boundary are treated using a Monte Carlo sampling approach: we place $N_{\rm samp}$ sampling points within the surrounding cells and count the number of points included in the particle.
The tracer quantities (e.g., gas mass, number of CRs) are deposited in cell $i$ as 
\begin{equation}
    q_i = \sum_a \frac{N_i^{(a)}}{N_{\rm samp}}q_{\rm trc}^{(a)}\frac{V_{\rm cell}}{V_{\rm trc}},
    \label{eq:deposit}
\end{equation}
where $V_{\rm mesh}$ is the volume of the mesh cell, and $N_i^{(a)}$ is the number of sampling points of the cell $i$ overlapping with the particle $a$. In this work, we use $N_{\rm samp} = 10^3$ points per cell.
This sampling method may overlook some particles when $V_{\rm trc}$ is smaller than $1/N_{\rm samp}$ of the cell volume.
To ensure the conservation of the tracer quantities, we calculate the sum of the sampled volume $\sum_a V_{\rm cell}N_i^{(a)}/{N_{\rm samp}}$ and deposit the remaining fraction, $V_{\rm trc}-\sum_a V_{\rm cell}N_i^{(a)}/{N_{\rm samp}}$, to the nearest grid cell and its neighbors with the CIC method.

\begin{figure*}
\includegraphics[width=0.9\linewidth]{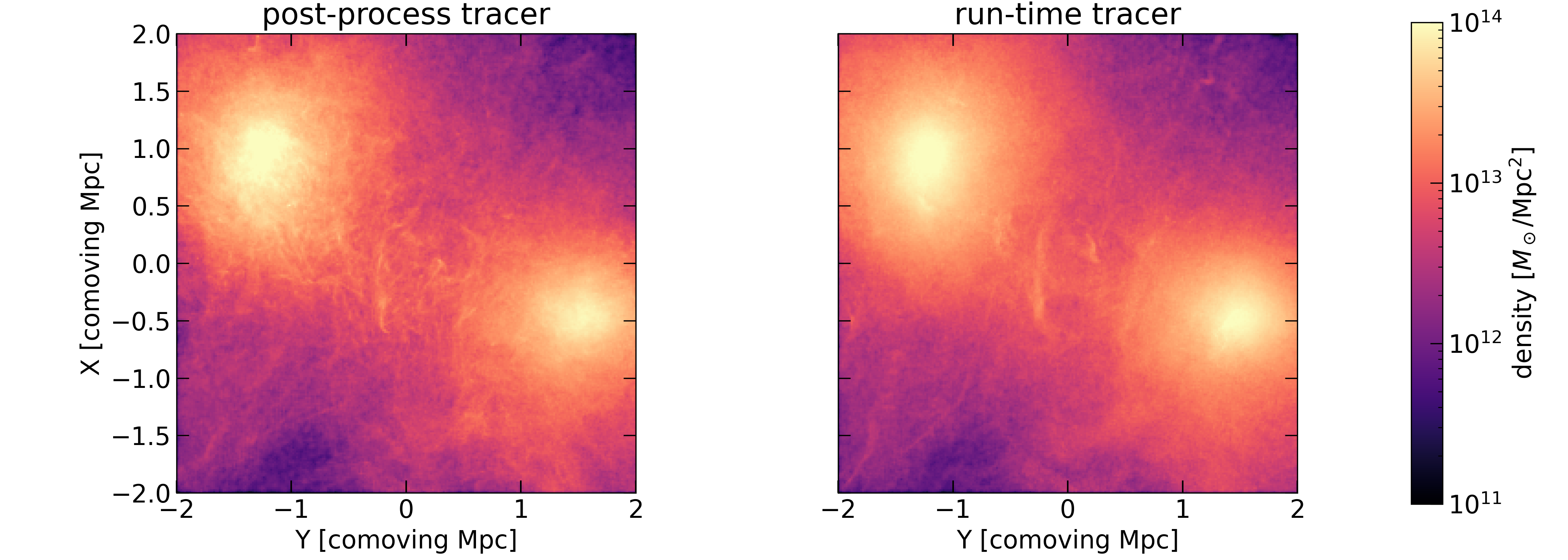}
\includegraphics[width=0.9\linewidth]{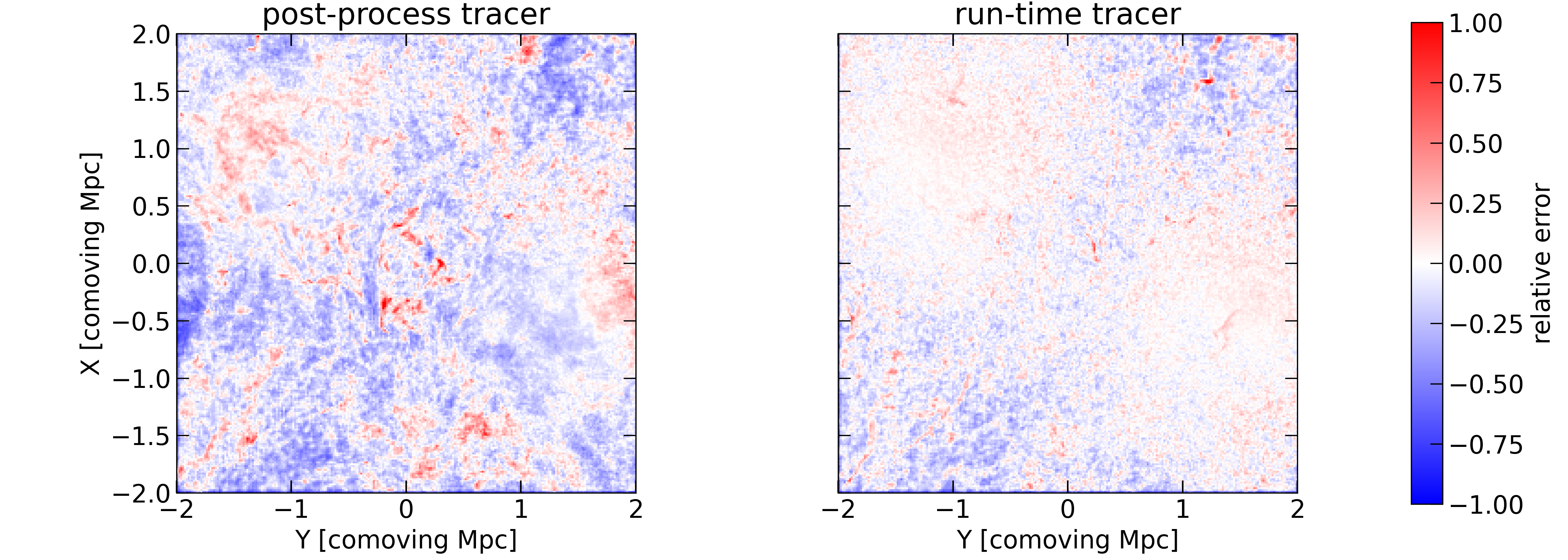}
  
    \caption{Comparison of projected density maps from tracer methods. Upper panels: projected density map calculated with the tracers. The left and right panels show the results for the post-process tracer method adopted in the previous works and the run-time tracer method developed in this work, respectively.
    Lower panels: the relative error to the original Enzo map.}
    \label{fig:compare_map}
\end{figure*}

\begin{figure*}
    \centering
    \includegraphics[width=0.9\linewidth]{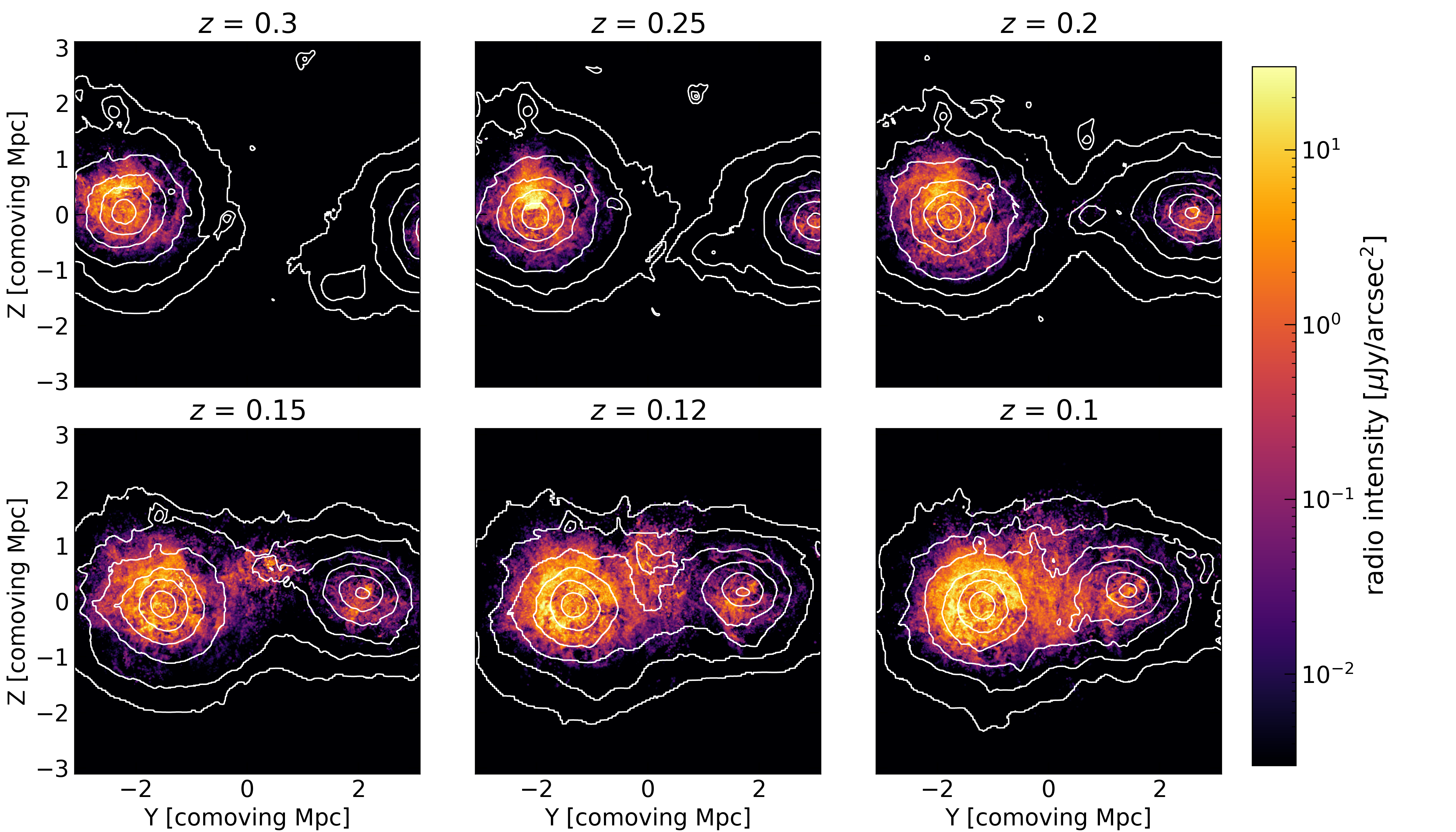}
    \caption{
    Same as Fig.~\ref{fig:bridge_radio_map_z}, but projection along X axis of the simulation.
    }
    \label{fig:bridge_radio_map_x}
\end{figure*}

\section{Quality of the tracer method}\label{app:trc_map}
In this section, we assess the quality of the tracer method by comparing the gas density maps reconstructed from the tracer data with those obtained directly from the original Enzo outputs.
As explained in Appendix~\ref{app:trc_generation}, the tracer particles are generated so as to reproduce the gas density distribution at the generation epoch ($z = 1$).
In general, the difference between the Enzo grid data and the tracer-based mass distribution is smallest at the generation epoch and gradually increases with time.
\par

We compare our run-time tracer method with the post-process tracers used in previous studies in terms of how accurately they recover the gas density distribution at $z = 0.1$.
To this end, we conduct a simple test of the old post-process tracer method. In this test, we use the snapshot of the run-time tracers at $z = 0.5$ as an initial condition for the post-process tracers.
Thus, the number of post-process tracers is the same as that of the run-time tracers, i.e., $N_{\rm trc} = 1.5\times10^7$.
\par

Using Enzo snapshots produced every $\Delta z = 0.01$, we construct the velocity field on a uniform grid with a spatial resolution of 16 kpc.
The uniform grid covers the central (10 Mpc)$^3$ region of the simulation.
We then evolve the post-process tracers between two consecutive snapshots by dividing the interval into 10 sub-steps.
At each sub-step, the tracer velocity is estimated by interpolating the gas velocity field both in space and time.
More specifically, the velocities at the positions of tracers are first obtained from the two snapshots using the cloud-in-cell (CIC) interpolation, and then linearly interpolated in time to obtain the velocity at each sub-step.
The displacement at each sub-step is calculated with the fourth-order Runge-Kutta method.
We take into account the velocity perturbation term of Eq.~(\ref{eq:delta_v_trc}).
\par

In Fig.~\ref{fig:compare_map}, we show the projected gas density maps in 4 Mpc box centered on the bridge region at $z = 0.1$. 
The left panels show the maps reconstructed from the old post-process tracers, while the right panels show those from our new run-time tracers.
We use the deposition method described in Appendix~\ref{app:trc_deposit} to create the tracer-based density maps at a spatial resolution of 16 kpc.
The bottom panels show the relative error with respect to the original Enzo density map (Fig.~\ref{fig:Enzo_map_z}).
The regions where the tracer-based map overestimates or underestimates the Enzo density are shown in red and blue, respectively.
\par

The relative error in the baryon density map is clearly smaller for the run-time tracers than for the post-process tracers.
The post-process tracer method produces more filamentary structures than the run-time tracer method.
Although such particle clustering is a well-known issue in velocity-field tracers \citep[e.g.,][]{Genel_2019}, this test demonstrates that our method alleviates this problem.
In the cluster and bridge regions, the relative error of the baryon density for the run-time tracers is roughly within $\pm 0.3$, whereas the post-process tracers show errors of up to $\pm 0.7$ in the bridge region.
Note that the quality of the post-process tracer method depends on several factors, including the mass resolution, the initial spatial distribution of tracers, and the redshift interval of the Enzo snapshots used for interpolation.

\section{Projection along X axis \label{app:LOS_z}}

Fig.~\ref{fig:bridge_radio_map_x} shows the maps of 140 MHz radio emission and gas density projected along X axis of the simulation. 
The Mpc-sized radio bridge can also be seen in this projection at $z \lesssim 0.15$.
As seen in the projection along Z axis (Fig.~\ref{fig:Enzo_map_z}), there is a small halo moving in the -X direction. 
This small halo overlaps with a part of the bridge region when the system is projected along X axis.
On the other hand, Fig.~\ref{fig:bridge_radio_map_x} shows that there are no significant substructures along the Z axis.
Since the merger axis is nearly aligned to the Y axis, we get qualitatively similar results for the one-dimensional profiles (Sect.~\ref{sec:bridge_profile}).
Fitting to the 140 MHz profile perpendicular to the bridge (almost along the Z axis) with a Gaussian function, we find the bridge width of 0.84 Mpc.

\end{appendix}

\end{document}